\newif \iffullversion \fullversiontrue
\newif \ifcomments \commentsfalse
\newif \iflaw \lawfalse

    \documentclass[letterpaper,twocolumn,10pt]{article}
    \usepackage{usenix}

\usepackage[dvipsnames]{xcolor}
\usepackage{booktabs}
\usepackage{makecell}
\usepackage{xurl}
\usepackage{caption}[table]
\usepackage{tabularx}
\usepackage{amsthm}
\usepackage{multirow}
\usepackage{enumitem}
\usepackage{framed}
\usepackage{mdframed}
\usepackage{tcolorbox}
\usepackage{amsfonts}
\usepackage{pdflscape}
\usepackage{tikz}
\usepackage{xcolor}
\usepackage{adjustbox}
\usetikzlibrary{backgrounds,fit,shapes.misc}
\usepackage{xspace}
\usepackage{makecell}

\newcommand{\bi}[1]{\textbf{\textit{#1}}}

\newcommand{\uid}[1]{\textcolor{NavyBlue}{\textsf{#1}}}
\newcommand{\uidleg}[1]{\hyperref[tbl:legislative-docs-summary]{\textcolor{NavyBlue}{\textsf{#1}}}}
\newcommand{\uidreg}[1]{\hyperref[tbl:regulatory-docs-summary]{\textcolor{NavyBlue}{\textsf{#1}}}}
\newcommand{\uidweb}[1]{\hyperref[fig:web-docs-summary]{\textcolor{NavyBlue}{\textsf{#1}}}}
\newcommand{\uidptn}[1]{\hyperref[itemize:patterns-summary]{\textcolor{NavyBlue}{\textsf{#1}}}}

\ifcomments
    \newcommand{\andres}[1]{\textsf{\small{\color{ForestGreen}{[Andrés: {#1}]}}}}
    \newcommand{\arka}[1]{\textsf{\small{\color{Blue}{[Arka: {#1}]}}}}
    \newcommand{\mir}[1]{\textsf{\small{\color{Brown}{[Miranda: {#1}]}}}}
    \newcommand{\sunoo}[1]{\textsf{\small{\color{Magenta}{[Sunoo: {#1}]}}}}
    \newcommand{\will}[1]{\textsf{\small{\color{Plum}{[Will: {#1}]}}}}
    \newcommand{\edit}[1]{\textsf{\small{\color{Blue}{#1}}}}
\else
    \newcommand{\andres}[1]{\ignorespaces}
    \newcommand{\arka}[1]{\ignorespaces}
    \newcommand{\mir}[1]{\ignorespaces}
    \newcommand{\sunoo}[1]{\ignorespaces}
    \newcommand{\will}[1]{\ignorespaces}
    \newcommand{\edit}[1]{\ignorespaces}
\fi

\newcommand{\totaldocuments}{49 }
\newcommand{\totalinitialcodingdocs}{31 }
\newcommand{\totalcodebookdevelopmentrounds}{6 }

\newcommand{\totalcodes}{56 }

\newlength{\saveparindent}
\newlength{\saveparskip}
\newcounter{ctr}

\newenvironment{newitemize}{%
\begin{list}{\mbox{}\hspace{5pt}$\bullet$\hfill}{\labelwidth=15pt%
\labelsep=5pt \leftmargin=20pt \topsep=3pt%
\setlength{\listparindent}{\saveparindent}%
\setlength{\parsep}{\saveparskip}%
\setlength{\itemsep}{3pt} }}{\end{list}}

\definecolor{enacted}{HTML}{117733}
\definecolor{rejected}{HTML}{882255}
\newcommand{\yesTbl}{\textcolor{enacted}{\checkmark}}
\newcommand{\noTbl}{\textcolor{rejected}{$\times$}}

\newtheorem{parameter}{Parameter}
\definecolor{paramColor}{HTML}{332288}
\newcommand{\param}[1]{\textbf{\textsf{\textcolor{paramColor}{P#1}}}}

\definecolor{patternColor}{HTML}{332288}
\newcommand{\pattern}[1]{\textbf{\textsf{\textcolor{patternColor}{L#1}}}}

\definecolor{keyTrendColor}{HTML}{332288}
\newtheorem{keyTrendEnv}{Key Trend}

\newenvironment{takeaway}
  {\begin{tcolorbox}[colframe=keyTrendColor]
   \begin{keyTrendEnv}}
  {\end{keyTrendEnv}
   \end{tcolorbox}}

\definecolor{openQuestionsColor}{HTML}{DDCC77}
\newtheorem{openQuestionEnv}{Open Question}

\newenvironment{openQuestion}
  {\begin{tcolorbox}[colframe=openQuestionsColor]
   \begin{openQuestionEnv}}
  {\end{openQuestionEnv}
   \end{tcolorbox}}

\newenvironment{researchQuestions}{
    \begin{enumerate}[label=\textbf{RQ\arabic*.}, ref=RQ\arabic*, leftmargin=1.25cm]
}{
    \end{enumerate}
}

\usepackage{titlesec}
\iflaw
    \renewcommand\thesection{\Roman{section}}
    
    \newcommand\onlythesubsection{\Alph{subsection}}
    \newcommand\onlythesubsubsection{\arabic{subsubsection}}
    
    \titleformat{\section}{\centering\normalfont\large\scshape}{\thesection.}{1em}{}
    \titleformat{\subsection}{\centering\normalfont\itshape}{\onlythesubsection.}{1em}{}
    \titleformat{\subsubsection}{\normalfont}{\onlythesubsubsection.}{1em}{}

\else

\fi

\makeatletter
\newcommand\Autoref[1]{\@first@ref#1,@}
\def\@throw@dot#1.#2@{#1}%
\def\@set@refname#1{%
    \edef\@tmp{\getrefbykeydefault{#1}{anchor}{}}%
    \xdef\@tmp{\expandafter\@throw@dot\@tmp.@}%
    \ltx@IfUndefined{\@tmp autorefnameplural}%
         {\def\@refname{\@nameuse{\@tmp autorefname}s}}%
         {\def\@refname{\@nameuse{\@tmp autorefnameplural}}}%
}
\def\@first@ref#1,#2{%
  \ifx#2@\autoref{#1}\let\@nextref\@gobble%
  \else%
    \@set@refname{#1}%
    \@refname~\ref{#1}%
    \let\@nextref\@next@ref%
  \fi%
  \@nextref#2%
}
\def\@next@ref#1,#2{%
   \ifx#2@ and~\ref{#1}\let\@nextref\@gobble%
   \else, \ref{#1}%
   \fi%
   \@nextref#2%
}
\makeatother

\newcommand{\secref}[1]{\Autoref{#1}}

\newcommand{\subh}[1]{\medskip \noindent \textbf{{#1}}}
\renewcommand{\paragraph}[1]{\subh{#1}}

    \title{``AI Watermarking'': Bridging Policy Discourse and Technical Capabilities}

    \author{
        Andrés Fábrega \\ Cornell University
        \and
        Arkaprabha Bhattacharya \\ Cornell University
        \and
        Miranda Christ \\ Columbia University
        \and
        Sunoo Park \\ New York University
    }

\date{}

\begin{document}

\maketitle

\begin{abstract}
\iflaw
\else
    The widespread deployment of generative artificial intelligence (AI) models has raised serious concerns about the proliferation of AI-generated content. This has led to a surge of interest in, and demand for, reliable tracking and detection mechanisms for content that is AI-generated, such as watermarking, metadata tagging, content tagging, and more. The problem has captured the attention of policymakers as well as the popular media, and a spate of recent bills in the US have sought to regulate the spread of AI content, and enforce or promote methods to track and label it. 
    
    This work performs a critical analysis of the policy discourse surrounding generative AI content transparency in the US and EU. Through a broad document selection methodology, we first collect a broad corpus of documents containing legislative language and policy-relevant discourse on the topic. We then analyze these through inductive coding, and leverage our coding to systematize these documents, identifying key patterns, gaps, and open questions. 
    We identify \emph{critical points of disconnect} between policy and technological capabilities and practice, and we highlight and discuss \emph{potential ambiguities and pitfalls} raised by the trends in our corpus. 
    
\fi
\end{abstract} 

\section{Introduction}\label{sec:intro}
Recent advances in generative artificial intelligence (AI) have made sophisticated AI models widely accessible to billions of users worldwide~\cite{openaiusers, geminiusers}. While these developments bring many benefits—like increased productivity~\cite{dell2023navigating,acemoglu2025simple}, enhanced automation~\cite{eloundou2024gpts,acemoglu2025simple}, and new creative outlets~\cite{lee2024empirical}—they also present security-relevant societal challenges, such as the proliferation of mis- and disinformation~\cite{Hate2024Double}, facilitation of fraud and scams~\cite{ai-scams}, plagiarism and misattribution~\cite{carobene2024rising}, recursive model training~\cite{shumailov2024ai}, and more. 

Of course, humans can produce and spread deceptive content, defraud and scam, plagiarize, and so on---with or without AI. Yet lately, problems related to AI-generated content have drawn more intense public attention, and fueled a growing interest in \emph{distinguishing AI-generated content from human-authored content}---a problem seen by many as a central challenge to the safe deployment of AI.

The explosion of interest in ways to reliably track and detect AI content has engaged both the security and ML communities, leading to the development of various \emph{content transparency mechanisms}, such as perceptible and imperceptible\footnote{The machine learning community generally refers to watermarks as ``[in]visible'' rather than ``[im]perceptible'', but we use the latter as watermarks can be applied across visual and non-visual modalities (e.g., audio).} watermarks~\cite{zhao2025watermarking}, metadata tags~\cite{c2pa}, content tags~\cite{gamage2025labeling}, post-hoc detectors~\cite{mitchell2023detectgpt}, and more. 

These and related concepts are sometimes imprecisely bundled under the term ``AI watermarking,'' especially in broader public discourse. We use \emph{watermarking} to refer more narrowly to \emph{mechanisms for embedding detectable patterns in AI-generated content} (see~\secref{sec:defs}).\footnote{In the paper title and a few other places, we make an exception and use the term ``AI watermarking'' in scare quotes to reflect the broader, overloaded sense in which it appears in the wider discourse. \label{fn:scarequotes}}

The state of the art in content transparency has evolved significantly in recent years (see~\secref{sec:industry}). But for these mechanisms to have impact, technical solutions may need to be accompanied by effective policy that governs their use.

First, policy may need to drive the \emph{widespread adoption of content transparency tools}. Adding disclosures to AI content seems unlikely to align with model providers' business interests, who may see such measures as discouraging \iffullversion their tools' use\else use\fi.\footnote{For example, in 2024, OpenAI reportedly did not add watermarks to ChatGPT due to, among other reasons, possible decrease in usage~\cite{openai-watermark-no-deployment-2}. Scott Aaronson also discussed this decision in a talk at the Simons Institute~\cite{scott-aaronson-talk}.} Thus, absent external pressure, content transparency may take a backseat to other financial priorities.

Second, policy may catalyze the \emph{development of content transparency standards}, currently a nascent area. There are many approaches to tracking and detecting AI content; the current variety and lack of consensus hinder interoperability and usability.
Furthermore, standards can aid the development and adoption of best practices~\cite{harcourt2020global,weiser2001internet}.
\iffullversion
    \footnote{But see~\cite{choi2025nist} for a different perspective.}
\fi%

Third, \emph{no content transparency solution is a silver bullet}---all tools (both present and future ones~\cite{zhang2024watermarks}) face important technical limitations. Mitigating the spread of malicious AI content cannot be solved by purely technical means, and must be complemented by broader socio-technical measures. %
For example, one could raise awareness that videos are not foolproof evidence of what they depict.

Fourth, government action (or other initiatives) may facilitate \emph{independent assessment of the effectiveness of content transparency solutions}, which can be critical to ensuring that adopted solutions are effective and secure. Policy efforts may be a catalyst for such steps.

\medskip
``AI watermarking'' is now prominent in the public consciousness and mainstream media, e.g., with an explainer in the New York Times~\cite{nyt-watermarking-explainer}.
There has also been a wide-ranging discourse around content transparency policy and legislation in recent years, resulting in a plethora of proposed bills and other regulatory and policy documents.

So far,
there appears to be little consensus amongst policymakers on goals, concerns, and legislative approaches.
Keeping up with the policy discourse is challenging due to its fast pace and fragmented nature, especially for policy outsiders, including technologists.
Perhaps relatedly, we find that there are some notable disconnects between policy discourse and technological practice (\secref{sec:leg}--\ref{sec:reg}). 

There is thus a pressing need for synthesis and analysis: to condense and evaluate the broader landscape of content transparency policy to date; to improve shared understanding across disciplinary boundaries; and to extract key patterns, technical considerations, and points of possible controversy. 
These are the concerns that motivate our present work.

Our paper provides the first \emph{systematization and analysis of policy discourse on ``AI watermarking''}, with the aim of
answering the following four main research questions (in which we abbreviate ``content transparency'' to ``CT'').

\begin{researchQuestions}
    \item\label{rq:define} How does existing policy discourse define ``AI watermarking,'' CT technologies, and their use cases?
    \item\label{rq:trends} What are common trends and divergences among legislative and regulatory documents on CT?
    \item\label{rq:align} To what extent does policy discourse on CT align with technical capabilities and practice?
    \item\label{rq:open} What are the main open questions and areas of uncertainty on how CT mechanisms should be regulated? 
\end{researchQuestions}

To tackle these questions, we assemble a wide-ranging corpus of legislative and other regulatory documents, focused on the United States (US), and to a lesser extent, the European Union (EU). With an in-depth document selection methodology (see \secref{sec:coding}), we assemble a corpus of official policy instruments. We then use inductive coding to analyze the diverse documents we assembled. Our analysis identifies key patterns, gaps, and open questions from our corpus, as follows.

\paragraph{Summary of contributions.}

\begin{enumerate}
    \item We present an overarching view of \bi{definitions of content transparency and related terms} in our corpus, noting ambiguities and inconsistencies
        (\secref{sec:defs}) 
    (\ref{rq:define}).
    \item We \bi{systematize 43 legislative documents}, resulting in a \bi{6-parameter analytical framework} that facilitates analysis and comparison across bills 
    (\secref{sec:leg})
    (\ref{rq:trends}).
    \item We distill \bi{10 key trends in content transparency legislation}, and examine how these relate to technical capabilities in practice 
    (\secref{sec:leg})
    (\ref{rq:trends}--\ref{rq:open}). 
    \item We curate and discuss \bi{10 open questions related to areas of uncertainty} in content transparency regulation, raised either in our corpus or by our analysis (\secref{sec:leg,sec:reg})
    (\ref{rq:trends}--\ref{rq:open}). 
    \item Our systematic analysis reveals \bi{issues absent or unaddressed in content transparency legislation, which are raised elsewhere} in our corpus, of which we provide extended discussion
        (\secref{sec:leg,sec:reg})
    (\ref{rq:trends}--\ref{rq:open}).
    \item Based on all of the above, we offer a non-exhaustive list of \bi{5 concrete recommendations} for the future development of AI content transparency policy
    (\secref{sec:recommendations})
    (\ref{rq:open}). 
\end{enumerate}

\iffullversion\else
    \paragraph{Significance for the security community.}
    We explore the (mis)alignment between technical guarantees and policy discourse on ``AI watermarking,'' a topic that overlaps security, cryptography, and AI. While our contribution is multi-disciplinary, its strongest ties are to the security community.

    First, the policy discourse we study seeks to address harms related to security implications of AI use with or without content transparency. Prior research has demonstrated that generative AI can heighten security-related risks in diverse contexts (e.g., facilitating phishing~\cite{kumar2024phish}, abuse~\cite{gibson2024analyzing}, and mis-/dis-information~\cite{shoaib2023deepfakes, desai2024gen}). A primary goal of content transparency (policy) is to address such risks, particularly in the presence of adversarial actors. Second, rigorous treatment of the technical guarantees of content transparency technologies has taken place largely in the security and cryptography communities, appearing prominently at these venues (\secref{sec:related} gives examples). 
    Third, our work illuminates points of disconnect between existing policy discourse and such technical guarantees, aiming to highlight open questions and research directions for the security community, as well as to bridge the security, AI, and policy communities. %

\fi

\paragraph{Roadmap.}
\secref{sec:method} presents our methodology.
\secref{sec:defs} presents how documents in our corpus scope content transparency and related terms, and clarifies our intended definitions of such terms as used in this paper. 
\secref{sec:industry} overviews key industry developments, to provide context for what follows.
    Sections~\ref{sec:leg} and \ref{sec:reg} respectively present our systematization and analysis of legislative and other regulatory documents.
\secref{sec:recommendations} presents our recommendations for future content transparency policy. \secref{sec:related} covers further related work, and \secref{sec:conclusion} concludes.

\section{Methodology}\label{sec:coding}\label{sec:method}
We collected \totaldocuments documents of two types which we call \emph{legislative} and \emph{other policy} documents, and examined them through inductive coding and thematic analysis as detailed below.

\subsection{Scope and types of documents} 
The first step in our coding process consisted of determining the \emph{types} of documents we would analyze. Our focus is on two types of documents that embody content transparency policy discourse: legislative documents and other policy documents. \emph{Legislative documents} are those involved in the lawmaking process; the two types we examine are (1) bills,\footnote{Bills are typically referred to by identifiers; for example, in the US, these correspond to a prefix denoting the chamber of Congress were the bill was introduced, followed by a counter denoting their order.} which are proposals for new laws, and (2) enacted laws. 
By \emph{other policy documents}, we refer to other documentation from government bodies that may guide and complement legislation. Examples of such documents include official guidance from standardization agencies, executive orders, briefings from working groups, etc. Our focus is on the US and US states (primarily) and the EU. 
    
Most of our documents are from the US federal and state contexts;\footnote{Our research team is most familiar with governance in the US.} we also examine several EU documents.\footnote{We include only EU-level documents, not member state documents.}

\subsection{Document selection} 
We built our document set by identifying relevant document repositories for each type of document, and performing keyword search on each of these. 

\paragraph{Timeframe.}
We systematically examined documents published between January 1, 2022 and January 3, 2025, starting when AI-based content generation became popular and widely accessible following the release of products such as ChatGPT~\cite{introducing-chatgpt}, and ending with the conclusion of legislative session of the 118th Congress about three years later~\cite{congress-conclusion}.

\paragraph{Sources of legislative documents.}\label{sec:A-list-doc-sources} 
Our scope for legislative documents in the US consisted of all proposed bills in a 2-year window across 51 legislatures (all states and US Congress).
We drew legislative documents from the following sources: the Brennan Center's Artificial Intelligence Legislation Tracker\footnote{\url{https://www.brennancenter.org/our-work/research-reports/artificial-intelligence-legislation-tracker}}; the National Conference of State Legislatures' Artificial Intelligence 2024 Legislation database\footnote{\url{https://www.ncsl.org/technology-and-communication/artificial-intelligence-2024-legislation}}; Public Citizen's State Legislation on Deepfakes in Elections tracker\footnote{\url{https://www.citizen.org/article/tracker-legislation-on-deepfakes-in-elections}}; and official European Parliament and Council Repositories.

\paragraph{Sources of other policy documents.}
Similar tracker databases do not exist for other policy documents broadly, so we referred to prominent sources that the research team was aware are active in policy discussion on emerging technologies, and references therein.
We sourced other policy documents from the following sources by searching their official websites and publications: the European Parliament; the European Council; the European Union Agency for Cybersecurity (ENISA); the US Cybersecurity and Infrastructure Security Agency (CISA); the US Federal Communications Commission (FCC); the US Federal Trade Commission (FTC); the US National Institute of Standards and Technology (NIST); and the US White House.
We did not include other policy documents at the US state level.

\paragraph{Selecting relevant documents from these sources.}
\iffullversion
    We searched the above sources 
    for relevant documents 
    using keywords. 
\fi
Starting with ``AI watermarking'' and ``AI watermarking policy,'' we refined and expanded our search terms iteratively, based on reading responsive documents.
\iffullversion
    Since one goal of this study is to characterize how documents define watermarking and its associated use cases, we kept our keywords general. 
     The full list of search terms used in our document selection process were: \textit{Watermarking, AI Watermark, AI Labeling, Watermark, Watermarking, Provenance, Generative Models, Deepfakes, Disclosure, AI Disclosures, Synthetic Content, AI, Synthetic}.

\else
    Appendix~\ref{sec:A-keywords} gives a full list of search terms and more detail on our search procedure.
\fi

This initial keyword search resulted in 50 total documents across both document categories. Two researchers then verified these documents' relevancy via separate reads, and excluded one of the documents as not relevant.\footnote{This document mentioned ``watermarking'' in the context of discussing an environmental concern.} Our selection process thus yielded a dataset of \totaldocuments total documents for analysis, consisting of 43 legislative documents (36 rejected bills and 7 enacted laws) and 6 other policy documents. We provide a public online repository of our documents.\cite{systematizationRepo}
Full lists of our collected documents are available in Tables \ref{tbl:legislative-docs-summary} and \ref{tbl:regulatory-docs-summary} in \secref{sec:leg,sec:reg} respectively. 
\bi{Throughout this Article, documents may be referenced by their unique identifiers in these tables.}

\subsection{Analysis}\label{sec:method:analysis}

\paragraph{Qualitative coding and analysis.} 
We next used \emph{inductive coding}~\cite{saldana2014thinking}, an iterative, bottom-up process to extract innate themes from texts, to evaluate our \totaldocuments documents.
Coding was performed by two members of the research team. First, the researchers performed an open-coding of a sample of \totalinitialcodingdocs documents over \totalcodebookdevelopmentrounds rounds, convening frequently to discuss common codes, disagreements, and any new codes identified in the process. This resulted in a codebook with \totalcodes lower level codes across 13 broader themes, summarized in \secref{tab:coding}.
Our researchers achieved a Cohen's Kappa score of 0.86, indicating very high agreement between researchers.\footnote{See \cite{mchugh2012interrater}.} 
{\centering
\begin{table}[h!]
\resizebox{\linewidth}{!}{
\begin{tabular}{p{0.2\linewidth}p{0.4\linewidth}p{0.5\linewidth}}
\toprule
\textbf{Code category} & \textbf{Definition} & \textbf{Example low level codes} \\ \midrule
Category & Document type & Bill (passed into law), Bill (rejected), Web article\\
Author category & Background of author  & Government official, Government Entity, Academic expert \\
Broader primitive & How do documents describe content transparency mechanisms and their hierarchy? & Focus on specific examples, Acknowledge a broader primitive \\
Use cases & Describes the set of use cases the document mentions with respect to watermarks. & Misinformation, NCII, Deepfakes,  Election Integrity\\
Mechanisms & Describes the set of mechanisms the document specifically mentions & Invisible watermarks, Cryptographic/Digital Signatures, Metadata tagging \\
Mechanism presentation & How documents describe the relationship of content transparency mechanisms they mention. & Examples, exhaustive list \\
Coordination with model developers & How does the document describe coordination with model developers for adding content transparency? & Does not discuss, Require model developer, Not applicable \\
Detection process & How does the document describe the detection process of a mechanism? & Acknowledges detection process, Provides implementation details, Does not acknowledge, Not applicable \\
Distinguishing verifiers and tool providers & How does the document describe the relationship between verifiers and tool providers? & Acknowledges that verifiers differ from providers, Does not address \\
Trust & Who are verifiers required to trust with content transparency mechanisms? & Not mentioned, model providers, detection-tool providers \\
Properties mentioned & The properties of mechanisms mentioned in document. & Adversarial Robustness, Honest Robustness, Quality, Completeness, Unforgeability, Soundness \\
Calls to action & Calls to action made by the document. & Model providers must add watermarks, No person should remove watermarks, Model providers should provide detection tools \\
Assertiveness & A measure of the assertiveness in the document's call to action. & Suggestion, strong suggestion, legal mandate, N/A \\
\bottomrule
\end{tabular}
}
\caption{The 13 high level coding categories and example low level codes for each.}
\label{tab:coding}
\end{table}
}

The two researchers then coded the remaining documents independently, reconvening as needed to discuss any emerging themes or points of analysis. Following best practices to ensure coding quality remained high, both researchers took breaks between coding rounds, and spot-checked codes throughout the analysis.

\paragraph{Thematic analysis.} After coding, we performed a thematic analysis~\cite{fereday2006demonstrating} of the resulting codes. We thus identify the patterns, key trends, open questions, and points of divergence elaborated in 
\secref{sec:leg,sec:reg}.

\subsection{Limitations of Methodology}

\iffullversion
    \paragraph{Jurisdictional limitations.}
    There have been many impactful policy developments in other jurisdictions outside the US and EU, perhaps most notably from China's Cyberspace Administration~\cite{china-ai-provisions}. Other examples include various Latin American countries~\cite{otherjurisdictions,otherjurisdictions2} and South Korea~\cite{sk-ai-provisions}. While acknowledging the importance of this global context, we limit our focus to the US and EU based on our own background and expertise, their status as key players in AI regulation, and to keep our scope manageable. Broader and/or comparative studies would be interesting future work.
    
    \paragraph{Stakeholders.}
    Our research focuses on policy-related discourse. The views of some stakeholders in AI watermarking may not be well represented in our work, whether because they are underrepresented in policy discussions, or because they are not policymakers, legislators, or otherwise involved in policy discourse. Our research focus is \emph{not} on laypeople's perceptions, although these may be incidentally represented in some of our documents.
\else
    \paragraph{Scope limitations.}
    (1) There have been many impactful policy developments in other jurisdictions outside the US and EU (e.g.,~\cite{china-ai-provisions,otherjurisdictions2,sk-ai-provisions}). 
    While acknowledging the importance of this global context, we limit our focus to the US and EU based on our own background and expertise, their status as key players in AI regulation, and to keep our scope manageable.
    (2) We focus on policy-related discourse. As such, the views of any stakeholders in AI watermarking who are are underrepresented in policy discussions will not be well represented in our work.
\fi

    \section{Key Concepts and Terminology}
\label{sec:defs}

Terminology around AI and content transparency is used inconsistently. Just agreeing on a definition of AI within a given context can be a major practical challenge~\cite{carnegie2022_ai_definition}, and no consistent scoping has been established by prior work.
This section overviews how watermarking terminology is used in our corpus, and also clarifies how these terms are used in this paper.

\subsection{Content}

\paragraph{Artificial intelligence.} By \emph{artificial intelligence (AI)}, we mean systems that can achieve human-assigned objectives by making choices based on data. By \emph{generative AI} (GenAI), we mean the subset of AI systems that generate or derive content based on data. 

Our corpus of legislative and other policy documents vary widely in defining AI. Six bills failed to define AI, and one bill, A 7904, defers to a definition by the New York State Board of Elections. Of the remaining 41, some scope AI to 
`technology or tools that use predictive algorithms to create new content, including audio, code, images, text, simulations or videos' (S 1044), which practitioners and academics would more likely refer to as \emph{generative artificial intelligence} (GenAI), a subcategory of AI. Others precisely use the term GenAI to refer to this kind of technology.

\iffullversion
H 5450 offers a broader view, defining AI as ``a machine-based system that (A) can, for a given set of human-defined objectives, make predictions, recommendations or decisions influencing real or virtual environments, and (B) uses machine and human-based inputs to (i) perceive real and virtual environments, (ii) abstract such perceptions into models through analysis in an automated manner, and (iii) formulate options for information or action through model inference''. 
\fi

\paragraph{Model.} By \emph{model}, we mean a system that produces human-interpretable \emph{outputs} based on \emph{inputs}.

Only a few documents explicitly discuss GenAI models, and some mention the concept in passing without defining it. For example, A 3050 mentions models while defining AI: ``AI includes various subfields, including, but not limited to, machine learning, natural language processing, and large language models.'' 

Others add specificity. For example, S 3312 defines a ``data model'' as : ``a mathematical, economic, or statistical representation of a system or process used to assist in making calculations and predictions, including through the use of algorithms, computer programs, or artificial intelligence systems...''.

\paragraph{Model provider.} 
By \emph{model provider}, we mean any entity that develops a model and/or directly allows or intermediates other entities' interaction with a model.  This interaction may be black-box, i.e., supplying inputs and receiving outputs; or white-box, i.e., having direct access to (parts of) the model, such as its weights, training data, etc. The model may be accessible to the broader public or to just a set of entities (for example, paying customers).

Some documents use terms that are broader than ours. For example, A 3050 defines ``AI-generating entity'' as ``an entity, including a business, that generates, creates, or otherwise produces AI-generated materials``. 
Others distinguish specific kinds of model providers, such as AB 3211, which explains that ``Generative AI provider” or ``GenAI provider” means an organization or individual that creates, codes, substantially modifies, or otherwise produces a generative AI system that is made publicly available for use by a California resident, regardless of whether the terms of that use include compensation.''

\paragraph{AI-generated content.} By \emph{AI-generated content}, we mean any media that was generated entirely using or includes any direct portions of generated content from GenAI tools.

While content covered by bills and regulation vary (as we will see in Sections \ref{sec:leg} and \ref{sec:reg}), the focus of this work is particularly on AI-generated content, which most bills define in some way. Some, like H 5450 only consider deceptive content as in scope: ``"Deceptive synthetic media" means any image, audio or video of an individual, ... which ... a reasonable person would believe depicts the appearance, speech or conduct of such individual when such individual did not in fact appear as depicted or engage in such speech or conduct, and ... was generated, in whole or in part, through the use of artificial intelligence or other means''. 

Others focus on whether content is ``synthetic'' to decide whether it is in-scope, such as A 7106, which pertains to any content ``which was produced by or includes any synthetic media''. 
Furthermore, while some documents do not delineate the level of AI use in generating content, bills like S 88 from North Carolina provide some scoping of the level of AI use, saying relevant content involves anything ``created in whole or in part with the use of generative artificial intelligence''. Documents are similarly vague about the specificity of AI use, as we explore later.

\paragraph{Metadata.}
 By \emph{metadata}, we mean structural or descriptive information about content. Mentions of metadata in bills often refer to \emph{44 U.S.C. \S\ 3502}, which defines metadata as ``structural or descriptive information about data such as content, format, source, rights, accuracy, provenance, frequency, periodicity, granularity, publisher or responsible party, contact information, method of collection, and other descriptions;'' 

\subsection{Detection and Technologies}

\paragraph{Disclosure.} The term ``disclosure'' is ambiguous both in our usage and within the corpus we analyze.
``Disclosure'' may have the connotation of public communication.
However, some content transparency mechanisms, such as invisible watermarks, are non-public in that they require a secret key for detection, so a narrow definition of ``disclosure'' may not include them.
Following many works in our corpus, we use a broad definition that includes public and non-public transparency mechanisms. In particular, by \emph{disclosure}, we mean processes for indicating the use of GenAI in content, both for self-disclosures and for model provider disclosures. Disclosures can be \emph{visible}, which are noticeable and interpretable by human perception of the content as typically exhibited; or \emph{invisible}, which are difficult to notice or interpret by human perception of the content as typically exhibited.

Disclosure is a frequently appearing term in legislation and other policy documents, used in a wide array of ways. For example, disclosure is often used to describe the \emph{process} of indicating that content was generated or modified using GenAI. In some cases, this refers to self-disclosure, where \emph{users} denote their published GenAI content as such (using, for example, explicit labels alongside content), as seen in, for example, HR 5586 by Congress. In other cases, this refers to provider disclosures, where model \emph{providers} enhances their output to signal that they were produced by GenAI.

On the other hand, disclosure is also used as the \emph{identifying information} itself. For instance, S 942 from California references ``manifest'' disclosure, which is described as ``easily perceived'', and ``latent'' disclosure, which is described as ``present, but not manifest''.

\paragraph{Content provenance \& provenance tracking.} In contrast, ``content provenance'' stands out as a term that is remarkably consistently used across all documents in our corpus. Most documents describe content provenance as information embedded into content or its metadata, for the purpose of identifying its authenticity, origin, and history of modification. We adopt this common definition for our work. In particular, we adopt the definition in S 942 from California: ``data that is embedded into digital content, or that is included in the digital content's metadata, for the purpose of verifying the digital content's authenticity, origin, or history of modification.'' We then define \emph{provenance tracking} as the process of tracking the history of generated content.

\paragraph{AI watermarking.} Overloading of the term ``watermarking'' is a source of confusion and friction in policy discourse. We find that the term is often used ambiguously and to mean different things in different contexts. For example, it is sometimes used as a specific subclass of mechanisms (e.g. invisible watermarks), as an umbrella term for multiple mechanisms (e.g., perceptible and imperceptible watermarks, or invisible watermarks and metadata tagging), or even as a broader term to capture all mechanisms for detecting and tracking the provenance of content. See Key Trend~\ref{kt:watermark-ambigous} in \secref{subsec:leg-takeaways} for further discussion.

\begin{figure}[t]
\centering
\begin{tikzpicture}[transform shape]
  \definecolor{WatermarkColor}{HTML}{009E73}
  \definecolor{InvisibleColor}{HTML}{D55E00}
  \definecolor{VisibleColor}{HTML}{CC79A7}

  \begin{scope}[blend mode=overlay]
    \fill[WatermarkColor, opacity=0.4] (0,0) circle (1.6);
    \fill[InvisibleColor, opacity=0.4] (1.5,0) circle (1.6);
    \fill[VisibleColor, opacity=0.4] (0.75,-1.4) circle (1.6);
  \end{scope}

  \draw[thick, WatermarkColor] (0,0) circle (1.6);
  \draw[thick, InvisibleColor] (1.5,0) circle (1.6);
  \draw[thick, VisibleColor] (0.75,-1.4) circle (1.6);

  \node at (-0.7,2) {\textbf{Watermarks}};
  \node at (2.7,2) {\textbf{Invisible disclosures}};
  \node at (0.75,-3.4) {\textbf{Visible disclosures}};

  \node at (0.75,0.6) {\small IW}; %
  \node[align=center] at (2.3,0.1) {\small Metadata \\ tags}; %
  \node at (-0.2,-1) {\small PW}; %
  \node at (0.75,-2.1) {\small Explicit labels}; %

  \node[
  draw=black,
  thick,
  rounded corners=8pt,
  inner sep=6pt,
  label={[align=center,yshift=2pt]above:\textbf{Content transparency mechanisms}},
  fit={( -2.7,2.3 ) ( 4.2,-3.8 )}
] {};

\end{tikzpicture}
\caption{Relationship between example terms that appear in content transparency policy. PW and IW stand for perceptible and imperceptible watermarks, respectively.}\label{fig:technologies}
\end{figure}
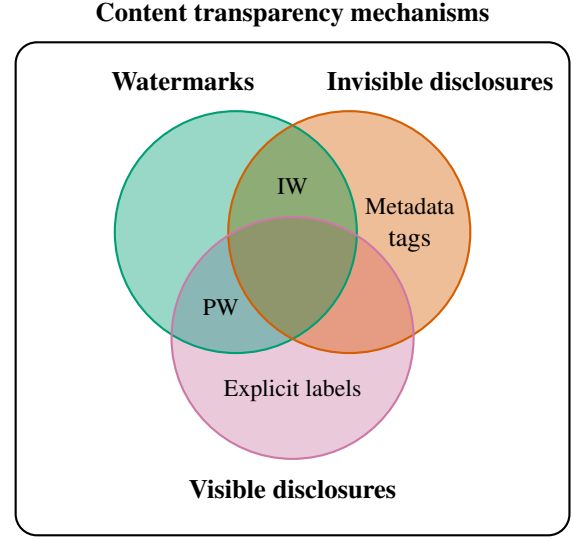

As shown in Figure~\ref{fig:technologies}, in this paper, we use \emph{content transparency mechanisms} to refer to the entire space of content provenance technologies. Concretely, following NIST's AI 100-4, we define it as ``methods for “documenting and accessing information about the origins and history of digital content''.
This is further subdivided into \emph{categories} of (more specifically-named) \emph{individual technologies}. In particular, by \emph{AI watermarking}, we mean two types of technologies: perceptible watermarking, which are watermarking mechanisms where embedded patterns are easily detectable by human perception. (e.g., Getty images or physical perturbations in content); and imperceptible watermarking, which are watermarking mechanisms where detection is performed by an algorithm, and the patterns are not easily
noticeable to human perception. This aligns closely to academic and practitioner views of these technologies. When documents discussed AI watermarking, they often referred to imperceptible watermarks. At its simplest, AI watermarking was described as a way to mark content as belonging to some original entity. 
For example, a CISA document titled \emph{Risk in Focus: Generative A.I. and the 2024 Election Cycle} defines AI watermarking as a way to  ``mark your content as verifiably originating from you''.

Meanwhile, the White House Executive Order, \emph{Safe, Secure, and Trustworthy Development and Use of Artificial Intelligence}, defines AI watermarking as ``the act of embedding information, which is typically difficult to remove, into outputs created by AI—including into outputs such as photos, videos, audio clips, or text for the purposes of verifying the authenticity of the output or the identity or characteristics of its provenance, modifications, or conveyance.'' 

In Figure~\ref{fig:technologies}, we show a graphical representation of the relationship between some example prominent terms that appear in content transparency policy.

    \section{Content Transparency in Industry}\label{sec:industry}

This section briefly overviews key industry developments in detecting AI content.
The practice of tagging content generated by production models is limited but growing, and currently applied most commonly to image models.

\paragraph{Metadata watermarks and C2PA.}
The Coalition for Content Provenance and Authenticity (C2PA)~\cite{c2pa}, founded by Adobe in 2019, establishes standards for attesting to the provenance of online content.
C2PA is an open organization collaborating with groups across industry and policy spaces.
Its steering committee includes several large organizations across different areas, including the Associated Press, Google, OpenAI, Nikon, and Sony.
Involved companies have already begun efforts to incorporate C2PA guidance into practice.  Meta~\cite{meta-watermark} and OpenAI~\cite{openai-watermark} have declared intent to incorporate C2PA-specified provenance data into their AI-generated images' metadata. 
Google has stated that they will use C2PA metadata to indicate which images in search results are AI-generated~\cite{google-watermark}. 
None has provided a concrete timeline.
Finally, a less prominent but notable alternative to C2PA is SEAL \cite{SEAL2025}, which enables cryptographic security of provenance metadata.

\paragraph{Perceptible watermarks.}
Several companies explicitly tag their AI-generated images with perceptible watermarks. 
For example, Meta applies a perceptible watermark specifically on photorealistic images it generates~\cite{meta-watermark}, and
Snapchat overlays a small ghost icon on images generated by its AI~\cite{snapchat-watermark}.

\paragraph{Imperceptible watermarks.}
To our knowledge, as of September 2025, only Google~\cite{synthid}, Meta~\cite{meta-watermark}, and Amazon~\cite{aws-watermark} have deployed imperceptible watermarks. 
(OpenAI developed a text watermarking scheme but ultimately decided not to deploy it.\footnote{The problems they stated included concerns about even a low false positive rate, stigmatizing legitimate use of LLMs by non-native speakers, and the ease with which their watermark could be removed~\cite{openai-watermark-no-deployment}. There is speculation that users' aversion toward watermarks contributed to this decision; 30\% of surveyed users said they would no longer use ChatGPT if its outputs were watermarked~\cite{openai-watermark-no-deployment-2}.})
Google's SynthID watermark family applies to text, images, and audio; Meta has announced only that they watermark images. 
While Amazon offers a publicly accessible detection interface or API,\footnote{API stands for Application Programming Interface. An API is an interface that can be accessed programmatically; in this case, this means that computer programs can send inputs to Amazon's watermark detection software and receive corresponding outputs.} neither Google nor Meta makes a detector publicly available.
Although both Google and Meta have published papers on their watermark schemes and have released open-source code, it is not clear how closely these match their deployed schemes.

\paragraph{Classifier-based detection.}
A less common way of identifying AI-generated content is using a machine learning classifier trained for this task.
This approach is susceptible to a high false positive rate~\cite{elkhatat2023evaluating} and ease of avoidance~\cite{liang2023gpt}. 
\iffullversion
    OpenAI once deployed a detection classifier, which it later took down; some speculate these types of concerns led to its removal~\cite{openai-watermark-no-deployment-3}.
\fi

\section{Analysis of Legislative Documents}
\label{sec:leg}

\iffullversion
    We now proceed to our analysis of watermarking legislative documents, derived from our qualitative coding of \secref{sec:coding}. We first set the stage with a broad summary of the legislative documents, followed by our analysis.
\fi
 
    \paragraph{Overview of documents.}
    Our document selection process, described in \secref{sec:coding}, identified a total of 43 legislative documents in the US and EU contexts. Thirty-two documents are at the US state level, 
    spanning a total of 18 states: Pennsylvania (2), Wisconsin, Connecticut (2), Mississippi (2), New York (7), North Carolina (2), California (5), Colorado, Florida, Utah, Ohio, Illinois, Oklahoma, Tennessee, Vermont, Louisiana, Hawaii, and Nebraska; the numbers in parentheses indicate the number of documents in each state, if more than one. Ten documents are federal bills, evenly divided across both chambers of Congress.\footnote{We write ``Congress'' to mean the US Congress.} Lastly, we include the preeminent AI-related EU legislation, the EU AI Act. Table~\ref{fig:legislative-docs-summary} gives a list of all our legislative documents, including their unique bill identifiers, jurisdiction, and source URLs.\footnote{We explain what the three ``pattern'' columns refer to in \secref{subsec:patterns}.}

The documents' focuses vary considerably, including disclosure of AI usage in election advertisement, inclusion of watermarks by model providers on all outputs, criminalization of removal of disclosures, and more.

\iffullversion
    Only 7 of the 43 bills were enacted into law: \uidleg{S 942} and \uidleg{A 2355} in California, \uidleg{A 664} in Wisconsin, \uidleg{H 1147} in Colorado, \uidleg{S 1680} in Florida, \uidleg{S 131} in Utah, and the \uidleg{EU AI Act}. For uniformity of reference, we refer to both bills and enacted laws by their bill numbers, which are unique within the scope of this Article. From the rejected bills, most were rejected as a result of concluding legislative periods, which results in the automatic rejection of all pending bills (which can then be re-introduced in the next period, if desired).\footnote{One of the bills, S 97 in Louisiana, was passed by the state's legislative branch but was subsequently vetoed by the Governor, citing First Amendment concerns and the need for further study of GenAI~\cite{s-97-veto}.}
\fi

\iflaw\else\paragraph{Section roadmap.} \fi 
Next, in \secref{subsec:bills-framework}, we derive a \emph{six-parameter taxonomy} from our inductive coding, designed to identify core components of content transparency legislation, to support structured comparison and analysis. 
\iffullversion
    In \secref{subsec:archetypes}, we apply this taxonomy to all legislative documents in our corpus, to obtain a broad systematization of content transparency legislation to date. 
\fi
We leverage this taxonomy to derive \emph{3 legislative patterns} that summarize common types of provisions, in \secref{subsec:patterns}. Finally, in \secref{subsec:leg-takeaways}, we provide a detailed discussion of \emph{key trends} and \emph{open questions} that we recommend policymakers consider in future policy and regulation. While open questions were not omitted from all documents, they were rarely addressed; we mention the cases where they were discussed.

    {\centering
    \begin{table*}
    \begin{tabular}{ll|lcl|ccc}
     \toprule
     \textbf{Identifier}\footnotemark & \textbf{Citation} & \textbf{Jdx} & \textbf{Enacted} & \textbf{Source} & \textbf{L1} & \textbf{L2} & \textbf{L3}\\ 
     \midrule
        \uid{A 3050} & {Assemb. B. 3050, Cal. Leg., 2023-24 Reg. Sess. (Cal. 2024)} & CA & \noTbl & \href{https://custom.statenet.com/public/resources.cgi?id=ID:bill:CA2023000A3050&cuiq=93d84396-c63b-526a-b152-38b7f79b4cfd&client_md=e4f6fea4-27b4-5d41-b7d3-766fe52569f0}{URL} (\href{https://perma.cc/XT6G-3P82}{P}) & \noTbl & \yesTbl & \yesTbl \\
        \uid{A 2355} & \makecell[l]{Assemb. B. 2355, Cal. Leg., 2023-24 Reg. Sess., ch. 260, \\ 2024 Cal. Stat. 93} & CA & \yesTbl & \href{https://custom.statenet.com/public/resources.cgi?id=ID:bill:CA2023000A2355&cuiq=93d84396-c63b-526a-b152-38b7f79b4cfd&client_md=e4f6fea4-27b4-5d41-b7d3-766fe52569f0}{URL} (\href{https://perma.cc/CR9G-RQAU}{P}) & \yesTbl & \noTbl & \noTbl \\
        \uid{A 664} & {Assemb. B. 664, 2023-24 Leg., 2023 Wis. Act 123} & WI & \yesTbl & \href{https://custom.statenet.com/public/resources.cgi?id=ID:bill:WI2023000A664&cuiq=93d84396-c63b-526a-b152-38b7f79b4cfd&client_md=e4f6fea4-27b4-5d41-b7d3-766fe52569f0}{URL} (\href{https://perma.cc/U6TV-RLY2}{P}) & \yesTbl & \noTbl & \noTbl \\
        \uid{A 7106} & {Assemb. B. 7106, N.Y. Leg., 2023-24 Reg. Sess. (N.Y. 2023)} & NY & \noTbl & \href{https://custom.statenet.com/public/resources.cgi?id=ID:bill:NY2023000A7106&cuiq=93d84396-c63b-526a-b152-38b7f79b4cfd&client_md=e4f6fea4-27b4-5d41-b7d3-766fe52569f0}{URL} (\href{https://perma.cc/UF3K-28D3}{P}) & \yesTbl & \noTbl & \noTbl \\
        \uid{A 7904} & {Assemb. B. 7904, N.Y. Leg., 2023-24 Reg. Sess. (N.Y. 2023)} & NY & \noTbl & \href{https://custom.statenet.com/public/resources.cgi?id=ID:bill:NY2023000A7904&cuiq=93d84396-c63b-526a-b152-38b7f79b4cfd&client_md=e4f6fea4-27b4-5d41-b7d3-766fe52569f0}{URL} (\href{https://perma.cc/X4ZW-F6S2}{P}) & \yesTbl & \noTbl & \noTbl \\
        \uid{A 9028} & {Assemb. B. 9028, N.Y. Leg., 2023-24 Reg. Sess. (N.Y. 2024)} & NY & \noTbl & \href{https://custom.statenet.com/public/resources.cgi?id=ID:bill:NY2023000A9028&cuiq=93d84396-c63b-526a-b152-38b7f79b4cfd&client_md=e4f6fea4-27b4-5d41-b7d3-766fe52569f0}{URL} (\href{https://perma.cc/XAM3-AEKV}{P}) & \yesTbl & \noTbl & \noTbl \\
        \uid{AB 3211} & {Assemb. B. 3211, Cal. Leg., 2023-24 Reg. Sess. (Cal. 2024)} & CA & \noTbl & \href{https://leginfo.legislature.ca.gov/faces/billNavClient.xhtml?bill_id=202320240AB3211}{URL} (\href{https://perma.cc/LD44-TLP8}{P}) & \noTbl & \yesTbl & \noTbl \\
        \uid{EU AI Act} & \makecell[l]{Regulation (EU) 2024/1689 of the European Parliament \\ and of the Council of 13 June 2024 laying down \\ harmonised rules on artificial intelligence \\ (Artificial Intelligence Act), 2024 O.J. (L 127) 1)} & EU & \yesTbl & \href{https://artificialintelligenceact.eu/the-act/}{URL} (\href{https://perma.cc/44MS-X86P}{P}) & \yesTbl & \yesTbl & \yesTbl \\
        \uid{H 1147} & {H.B. 1147, 74th Gen. Assemb., 2d. Reg. Sess. (Colo. 2024)} & CO & \yesTbl & \href{https://custom.statenet.com/public/resources.cgi?id=ID:bill:CO2024000H1147&cuiq=93d84396-c63b-526a-b152-38b7f79b4cfd&client_md=e4f6fea4-27b4-5d41-b7d3-766fe52569f0}{URL} (\href{https://perma.cc/UD9J-VKK5}{P}) & \yesTbl & \noTbl & \noTbl \\
        \uid{H 1267} & {H.B. 1267, Miss. Leg., 2024 Reg. Sess. (Miss. 2024)} & MS & \noTbl & \href{https://custom.statenet.com/public/resources.cgi?id=ID:bill:MS2024000H1267&cuiq=93d84396-c63b-526a-b152-38b7f79b4cfd&client_md=e4f6fea4-27b4-5d41-b7d3-766fe52569f0}{URL} (\href{https://perma.cc/QS69-FZBV}{P}) & \yesTbl & \noTbl & \noTbl \\
        \uid{H 1734} & {H.B. 1734, 32d Leg., 2024 Reg. Sess. (Haw. 2024)} & HI & \noTbl & \href{https://custom.statenet.com/public/resources.cgi?id=ID:bill:HI2023000H1734&cuiq=93d84396-c63b-526a-b152-38b7f79b4cfd&client_md=e4f6fea4-27b4-5d41-b7d3-766fe52569f0}{URL} (\href{https://perma.cc/3V8C-L6EK}{P}) & \yesTbl & \noTbl & \noTbl \\
        \uid{H 2660}& {H.B. 2660, 2024 Gen. Assemb., 2023-24 Reg. Sess. (Pa. 2024)} & PA & \noTbl & \href{https://www.palegis.us/legislation/bills/2023/hb2660}{URL} (\href{https://perma.cc/KK4A-8SLY}{P}) & \yesTbl & \noTbl & \noTbl \\
        \uid{H 2707} & {H.B. 2707, 113th Gen. Assemb., Reg. Sess. (Tenn. 2024)} & TN & \noTbl & \href{https://custom.statenet.com/public/resources.cgi?id=ID:bill:TN2023000H2707&cuiq=93d84396-c63b-526a-b152-38b7f79b4cfd&client_md=e4f6fea4-27b4-5d41-b7d3-766fe52569f0}{URL} (\href{https://perma.cc/VSA2-RSRK}{P}) & \yesTbl & \noTbl & \noTbl \\
        \uid{H 3453} & {H.B. 3453, 59th Leg., 2d. Sess. (Okla. 2024)} & OK & \noTbl & \href{https://custom.statenet.com/public/resources.cgi?id=ID:bill:OK2023000H3453&cuiq=93d84396-c63b-526a-b152-38b7f79b4cfd&client_md=e4f6fea4-27b4-5d41-b7d3-766fe52569f0}{URL} (\href{https://perma.cc/9E89-HC2V}{P}) & \noTbl & \yesTbl & \noTbl \\
        \uid{H 5321} & {H.B. 5321, 103d. Gen. Assemb., Reg. Sess. (Ill. 2024)} & IL & \noTbl & \href{https://custom.statenet.com/public/resources.cgi?id=ID:bill:IL2023000H5321&cuiq=93d84396-c63b-526a-b152-38b7f79b4cfd&client_md=e4f6fea4-27b4-5d41-b7d3-766fe52569f0}{URL} (\href{https://perma.cc/YS34-R39M}{P}) & \noTbl & \yesTbl & \noTbl \\
        \uid{H 5450} & {H.B. 5450, 2024 Gen. Assemb., Feb. Sess. (Conn. 2024)} & CT & \noTbl & \href{https://custom.statenet.com/public/resources.cgi?id=ID:bill:CT2024000H5450&cuiq=93d84396-c63b-526a-b152-38b7f79b4cfd&client_md=e4f6fea4-27b4-5d41-b7d3-766fe52569f0}{URL} (\href{https://perma.cc/DM6Z-DTXN}{P}) & \yesTbl & \noTbl & \noTbl \\
        \uid{H 710} & {H.B. 710, 2024 Gen. Assemb., Reg. Sess. (Vt. 2024)} & VT & \noTbl & \href{https://custom.statenet.com/public/resources.cgi?id=ID:bill:VT2023000H710&cuiq=93d84396-c63b-526a-b152-38b7f79b4cfd&client_md=e4f6fea4-27b4-5d41-b7d3-766fe52569f0}{URL} (\href{https://perma.cc/JW3T-3LTD}{P}) & \noTbl & \yesTbl & \noTbl \\
        \uid{HR 3831} & {H.R. 3831, 118th Cong. (2023)} & US & \noTbl & \href{https://www.congress.gov/bill/118th-congress/house-bill/3831/text}{URL} (\href{https://perma.cc/T6S4-ERQE}{P}) & \noTbl & \yesTbl & \noTbl \\
        \uid{HR 5586} & {H.R. 5586, 118th Cong. (2023)} & US & \noTbl & \href{https://www.congress.gov/bill/118th-congress/house-bill/5586/text}{URL} (\href{https://perma.cc/4UD6-3E3Z}{P}) & \yesTbl & \noTbl & \noTbl \\
        \uid{HR 7766} & {H.R. 7766, 118th Cong. (2024)} & US & \noTbl & \href{https://www.congress.gov/bill/118th-congress/house-bill/7766/text}{URL} (\href{https://perma.cc/E3GE-X25B}{P}) & \noTbl & \yesTbl & \yesTbl \\
        \uid{HR 8668} & {H.R. 8668, 118th Cong. (2024)} & US & \noTbl & \href{https://www.congress.gov/bill/118th-congress/house-bill/8668/text}{URL} (\href{https://perma.cc/E5XN-3WV3}{P}) & \yesTbl & \noTbl & \noTbl \\
        \uid{HR 9042} & {H.R. 9042, 118th Cong. (2024)} & US & \noTbl & \href{https://www.congress.gov/bill/118th-congress/house-bill/9042/text}{URL} (\href{https://perma.cc/248J-CBD7}{P}) & \noTbl & \noTbl & \yesTbl \\
        \uid{L 1203} & {Legis. B. 1203, 108th Leg., 2d. Sess. (Neb. 2024)} & NE & \noTbl & \href{https://custom.statenet.com/public/resources.cgi?id=ID:bill:NE2023000L1203&cuiq=93d84396-c63b-526a-b152-38b7f79b4cfd&client_md=e4f6fea4-27b4-5d41-b7d3-766fe52569f0}{URL} (\href{https://perma.cc/3RDK-LWNV}{P}) & \yesTbl & \noTbl & \noTbl \\
        \uid{S 1044} & {S.B. 1044, 2024 Gen. Assemb., 2023-24 Reg. Sess. (Pa. 2024)} & PA & \noTbl & \href{https://custom.statenet.com/public/resources.cgi?id=ID:bill:PA2023000S1044&cuiq=93d84396-c63b-526a-b152-38b7f79b4cfd&client_md=e4f6fea4-27b4-5d41-b7d3-766fe52569f0}{URL} (\href{https://perma.cc/6CUQ-MQGH}{P}) & \yesTbl & \noTbl & \noTbl \\
        \uid{S 1235} & {S.B. 1235, Cal. Leg., 2023-24 Reg. Sess. (Cal. 2024)} & CA & \noTbl & \href{https://custom.statenet.com/public/resources.cgi?id=ID:bill:CA2023000S1235&cuiq=93d84396-c63b-526a-b152-38b7f79b4cfd&client_md=e4f6fea4-27b4-5d41-b7d3-766fe52569f0}{URL} (\href{https://perma.cc/A5ZS-FVAC}{P}) & \noTbl & \noTbl & \yesTbl \\
        \uid{S 131} & {S.B. 131, 2024 Leg., Gen. Sess. (Utah 2024)} & UT & \yesTbl & \href{https://custom.statenet.com/public/resources.cgi?id=ID:bill:UT2024000S131&cuiq=93d84396-c63b-526a-b152-38b7f79b4cfd&client_md=e4f6fea4-27b4-5d41-b7d3-766fe52569f0}{URL} (\href{https://perma.cc/XVP7-H4KZ}{P}) & \yesTbl & \noTbl & \noTbl \\
        \uid{S 1596} & {S. 1596, 118th Cong. (2023)} & US & \noTbl & \href{https://www.congress.gov/bill/118th-congress/senate-bill/1596/text}{URL} (\href{https://perma.cc/YRX2-DXSU}{P}) & \yesTbl & \noTbl & \noTbl \\
        \uid{S 1680} & \makecell[l]{S.B. 1680, Fla. Leg., 2024 Reg. Sess., \\ ch. 2024-118, 2024 Fla. Stat. 1} & FL & \yesTbl & \href{https://flsenate.gov/Session/Bill/2024/1680/?Tab=BillText}{URL} (\href{https://perma.cc/6RYW-XGFV}{P}) & \noTbl & \noTbl & \yesTbl \\
        \uid{S 217} & {S.B. 217, 135th Gen. Assemb., Reg. Sess. (Ohio 2024)} & OH & \noTbl & \href{https://custom.statenet.com/public/resources.cgi?id=ID:bill:OH2023000S217&cuiq=93d84396-c63b-526a-b152-38b7f79b4cfd&client_md=e4f6fea4-27b4-5d41-b7d3-766fe52569f0}{URL} (\href{https://perma.cc/MG2C-WRS5}{P}) & \noTbl & \yesTbl & \noTbl \\
        \uid{S 2423} & {S.B. 2423, Miss. Leg., 2024 Reg. Sess. (Miss. 2024)} & MS & \noTbl & \href{https://custom.statenet.com/public/resources.cgi?id=ID:bill:MS2024000S2423&cuiq=93d84396-c63b-526a-b152-38b7f79b4cfd&client_md=e4f6fea4-27b4-5d41-b7d3-766fe52569f0}{URL} (\href{https://perma.cc/9WK2-99Y4}{P}) & \yesTbl & \noTbl & \noTbl \\
        \uid{S 2691} & {S. 2691, 118th Cong. (2023)} & US & \noTbl & \href{https://www.congress.gov/bill/118th-congress/senate-bill/2691/text}{URL} (\href{https://perma.cc/KYA5-SP83}{P}) & \noTbl & \yesTbl & \yesTbl \\
        \uid{S 2765} & {S. 2765, 118th Cong. (2023)} & US & \noTbl & \href{https://www.congress.gov/bill/118th-congress/senate-bill/2765/text}{URL} (\href{https://perma.cc/H8GD-GZTM}{P}) & \noTbl & \yesTbl & \yesTbl \\
        \uid{S 2} & {Substitute B. 2, 2024 Gen. Assemb., Feb. Sess. (Conn. 2024)} & CT & \noTbl & \href{https://custom.statenet.com/public/resources.cgi?id=ID:bill:CT2024000S2&cuiq=93d84396-c63b-526a-b152-38b7f79b4cfd&client_md=e4f6fea4-27b4-5d41-b7d3-766fe52569f0}{URL} (\href{https://perma.cc/7N6B-23H8}{P}) & \noTbl & \yesTbl & \noTbl \\
        \uid{S 3312} & {S. 3312, 118th Cong. (2024)} & US & \noTbl & \href{https://www.congress.gov/bill/118th-congress/senate-bill/3312/text}{URL} (\href{https://perma.cc/457E-ASBN}{P}) & \noTbl & \noTbl & \yesTbl \\
        \uid{S 4674} & {S. 4674, 118th Cong. (2024)} & US & \noTbl & \href{https://www.congress.gov/bill/118th-congress/senate-bill/4674/text}{URL} (\href{https://perma.cc/2BWA-2JAW}{P}) & \noTbl & \yesTbl & \yesTbl \\
        \uid{S 6859} & {S.B. 6859, N.Y. Leg., 2023-24 Reg. Sess. (N.Y. 2023)} & NY & \noTbl & \href{https://custom.statenet.com/public/resources.cgi?id=ID:bill:NY2023000S6859&cuiq=93d84396-c63b-526a-b152-38b7f79b4cfd&client_md=e4f6fea4-27b4-5d41-b7d3-766fe52569f0}{URL} (\href{https://perma.cc/L6BF-JRBC}{P}) & \yesTbl & \noTbl & \noTbl \\
        \uid{S 7847} & {S.B. 7847, N.Y. Leg., 2023-24 Reg. Sess. (N.Y. 2023)} & NY & \noTbl & \href{https://custom.statenet.com/public/resources.cgi?id=ID:bill:NY2023000S7847&cuiq=93d84396-c63b-526a-b152-38b7f79b4cfd&client_md=e4f6fea4-27b4-5d41-b7d3-766fe52569f0}{URL} (\href{https://perma.cc/CE5N-CEML}{P}) & \yesTbl & \noTbl & \noTbl \\
        \uid{S 7922} & {S.B. 7922, N.Y. Leg., 2023-24 Reg. Sess. (N.Y. 2024)} & NY & \noTbl & \href{https://custom.statenet.com/public/resources.cgi?id=ID:bill:NY2023000S7922&cuiq=93d84396-c63b-526a-b152-38b7f79b4cfd&client_md=e4f6fea4-27b4-5d41-b7d3-766fe52569f0}{URL} (\href{https://perma.cc/B2BD-PJ6G}{P}) & \yesTbl & \noTbl & \noTbl \\
        \uid{S 88} & {S.B. 88, 2023 Gen. Assemb., 3d. Sess. (N.C. 2023)} & NC & \noTbl & \href{https://custom.statenet.com/public/resources.cgi?id=ID:bill:NC2023000S88&cuiq=93d84396-c63b-526a-b152-38b7f79b4cfd&client_md=e4f6fea4-27b4-5d41-b7d3-766fe52569f0}{URL} (\href{https://perma.cc/ZKG7-4Z7L}{P}) & \yesTbl & \noTbl & \noTbl \\
        \uid{S 880} & {S.B. 880, 2023 Gen. Assemb., 1st Sess. (N.C. 2023)} & NC & \noTbl & \href{https://custom.statenet.com/public/resources.cgi?id=ID:bill:NC2023000S880&cuiq=93d84396-c63b-526a-b152-38b7f79b4cfd&client_md=e4f6fea4-27b4-5d41-b7d3-766fe52569f0}{URL} (\href{https://perma.cc/TE46-5H3V}{P}) & \yesTbl & \noTbl & \noTbl \\
        \uid{S 942} & \makecell[l]{S.B. 942, Cal. Leg., 2023-24 Reg. Sess., ch. 291, \\ 2024 Cal. Stat. 91} & CA & \yesTbl & \href{https://custom.statenet.com/public/resources.cgi?id=ID:bill:CA2023000S942&cuiq=93d84396-c63b-526a-b152-38b7f79b4cfd&client_md=e4f6fea4-27b4-5d41-b7d3-766fe52569f0}{URL} (\href{https://perma.cc/H4NQ-H22B}{P}) & \noTbl & \yesTbl & \noTbl \\
        \uid{S 97} & {S.B. 97, 2024 Leg., Reg. Sess. (La. 2024)} & LA & \noTbl & \href{https://custom.statenet.com/public/resources.cgi?id=ID:bill:LA2024000S97&cuiq=93d84396-c63b-526a-b152-38b7f79b4cfd&client_md=e4f6fea4-27b4-5d41-b7d3-766fe52569f0}{URL} (\href{https://perma.cc/V7GM-K2NY}{P}) & \yesTbl & \noTbl & \noTbl \\
        \uid{S 9640} & {S.B. 9640, N.Y. Leg., 2023-24 Reg. Sess. (N.Y. 2024)} & NY & \noTbl & \href{https://custom.statenet.com/public/resources.cgi?id=ID:bill:NY2023000S9640&cuiq=93d84396-c63b-526a-b152-38b7f79b4cfd&client_md=e4f6fea4-27b4-5d41-b7d3-766fe52569f0}{URL} (\href{https://perma.cc/H73V-7WWF}{P}) & \yesTbl & \noTbl & \noTbl \\
        \midrule
        \multicolumn{5}{r|}{Totals} & 26 & 14 & 10 \\
     \bottomrule
\end{tabular}
\caption{\textbf{Full list of legislative documents.} ``Jdx'' stands for jurisdiction. Permalinks are linked in parentheses. Last three columns refer to the patterns in \secref{subsec:patterns}.}
\label{fig:legislative-docs-summary}
\label{tbl:legislative-docs-summary}
\end{table*}

\footnotetext{Bill numbering systems are separate in different jurisdictions, but it happens that we have no duplicate identifiers in our corpus. For brevity, we sometimes refer to bills just by their identifier without the jurisdiction.}
}

\begin{table}[t!]
    \centering
    \iflaw\else\footnotesize\fi
    \renewcommand{\arraystretch}{1}
    \begin{tabular}{llr}
    \toprule
      \textbf{Parameter} & \textbf{Parameter values} & \textbf{\# docs} \\ 
     \midrule
     \multirow{4}{*}{\shortstack[l]{\param{1} \\content scope}}  & Unqualified usage of AI tools & 17 \\
     & Qualified usage of AI tools (quantitative) & 8 \\
     & Qualified usage of AI tools (qualitative) & 15 \\
     & Not defined & 3 \\
    \midrule
     \multirow{3}{*}{\shortstack[l]{\param{2} \\use cases}}  & General & 20 \\
     & Specific & 23 \\
     & Exceptions & 6 \\
    \midrule
     \multirow{4}{*}{\shortstack[l]{\param{3} \\mandates}}  & Model providers must add disclosures & 14 \\
     & Model users must proactively disclosure & 26 \\
     & Disclosures must not be removed & 7 \\
     & Gov't agencies must establish standards & 10 \\
    \midrule
     \multirow{2}{*}{\shortstack[l]{\param{4} \\disc. mech.}}  & Broad & 38 \\
     & Specific & 5 \\
    \midrule
     \multirow{3}{*}{\shortstack[l]{\param{5} \\requirements on\\disc. mech.}}  & No requirements & 7 \\
     & Qualitative requirements & 26 \\
     & Specific requirements & 25 \\
    \midrule
     \multirow{3}{*}{\shortstack[l]{\param{6} \\penalties}}  & Monetary penalties & 13 \\
     & Violation of existing statute & 8 \\
     & No penalties mentioned & 22 \\
    \bottomrule
\end{tabular}
\caption{%
    \textbf{The six parameters of our taxonomy, or, ``anatomy of a content transparency bill''} (\secref{sec:anatomy}). A bill may have multiple values per parameter, although some values are mutually exclusive. Not all bills have values for all parameters.
}
\label{tbl:framework-summary}
\end{table}

\subsection{Anatomy of a Content Transparency Bill}\label{subsec:bills-framework}\label{sec:anatomy}

\iffullversion
    A central challenge in the analysis of legislative documents is that their details vary considerably, making it difficult to extract general conclusions. Therefore, the first step of our analysis consisted of leveraging the results of our coding process (described in \secref{sec:coding}) to derive a \emph{taxonomy} which systematizes the structure and content of all forty-three legislative documents, characterizing common patterns and identifying key differences. 
    
    To arrive at our taxonomy, which we present in this section, we performed inductive framework analysis~\cite{becker2007epistemological} over our resulting themes from \secref{sec:coding}. We observed clusters of themes emerging within legislative documents, which we assembled into \emph{six core parameters} that capture the high-level structure of watermarking bills, extracting their key components while abstracting away lower-level details. Individual documents can then be categorized in our taxonomy by the concrete parameter values applicable to them.
    
    Our taxonomy thus provides a structured lens to interpret and compare documents, and to distill general trends and takeaways. A framework of this type cannot---and is not designed to---capture all details of the individual documents; however, the big-picture view that our taxonomy offers may help identify relevant documents to examine in more depth.
\else
    A central challenge in analyzing legislative documents is that their details vary considerably, making it hard to extract general conclusions. Therefore, the first step of our analysis consisted of leveraging the results of our coding process (\secref{sec:coding}) to derive an \emph{analytical framework} which systematizes the structure and content of all forty-three legislative documents. During our thematic analysis, we noticed clusters of themes emerging within legislative documents, which we assembled into \emph{six core parameters} that capture the high-level structure of content transparency bills.
\fi

In a nutshell, the six parameters of our taxonomy are: (\param{1}) \emph{scope of content} covered by the bill; (\param{2}) \emph{use cases} the bill applies to; (\param{3}) \emph{mandates} of the bill; (\param{4}) acceptable \emph{disclosure mechanisms} that satisfy the mandates; (\param{5}) \emph{requirements} imposed on these mechanisms; and (\param{6}) \emph{penalties and consequences} for violations. 
\param{1}--\param{3} are relevant to every document in our corpus, whereas \param{4}--\param{6} are instantiated only for documents that discuss or regulate disclosures (all but four in our corpus). %
The parameters and their possible values are summarized in Table~\ref{tbl:framework-summary}, and detailed below. In \secref{subsec:leg-takeaways}, we discuss trends in the way in which legislative documents instantiate these parameters.

\begin{parameter}\label{param:content-scope}
What is the scope of AI-generated content?
\end{parameter}
The first parameter defines the \emph{type of content} that falls under the provisions of the bill. This corresponds to defining the content that is considered ``AI-generated''. Specifying this parameter can be broken down into two sub-questions: (1) \emph{how is ``AI'' defined?}, and (2) \emph{what content is covered by legislation, with respect to the AI definition in step (1)?} 
\iffullversion
    For both sub-parameters, a bill's definitions may be directly embedded in the body of the bill, or may be referenced from an external source (e.g., other AI-related bills or regulations).
\fi

Across our legislative corpus, we observed three ways this parameter is instantiated. First, many bills define their scope of covered content as determined by whether AI was used \emph{at all or not} to generate content. For example, common verbiage across these bills is ``generated in whole or in part with the use of artificial intelligence'' (e.g., \uidleg{S 2423} in Mississippi or \uidleg{A 7904} in New York). We refer to this parameter instantiation as the \bi{unqualified usage of AI tools}.

Other bills have more fine-grained scoping, defining covered content based on the \emph{degree} to which AI altered the original content. For example, such bills may cover content only if it is ``substantially altered'' by AI so that a ``reasonable observer'' could not discern whether the content is real or fake. Such qualifications may be specified quantitatively or qualitatively, as discussed in \secref{subsec:leg-takeaways}. We refer to this parameter instantiation as the \bi{qualified usage of AI tools}.

Lastly, some bills, such as \uidleg{HR 3831} in Congress, do \bi{not define} a scope of covered content. %

\begin{parameter}\label{param:use-cases}
What AI use cases is the bill applicable to?
\end{parameter}
The second parameter defines the \emph{contexts} and \emph{use cases} in which a bill regulates AI content. Together, \param{1} and~\param{2} define a bill's scope of applicability. 

We observed three broad patterns. Some bills are \bi{general}, applicable to any AI content irrespective of how and where it is used. Others focus on \bi{specific} use cases and contexts (e.g., election communications). Some bills have narrow scope, excluding use cases via \bi{exceptions} (e.g., news reporting).

\begin{parameter}\label{param:provisions}
    What is being mandated from whom? 
\end{parameter}
The third parameter defines the \emph{mandates} of the bill, i.e., the way in which AI content (as scoped in \param{1} and in the use cases from \param{2}) is regulated. At a high level, this parameter defines \emph{who} is responsible and \emph{what} is asked of this entity. 

We observed four values for this parameter: the first three relate to \emph{disclosures} and the last relates to \emph{standards}. Here, we just consider what sort of disclosure action is mandated from whom; \param{4} and \param{5} elaborate the nature of required disclosures.

First, some bills mandate that \emph{model providers} must \bi{add disclosures} to content produced by their tools. Such bills generally (but not always) specify criteria for covered model providers, such as determined by their number of users. Second, other bills mandate that model \emph{users} must \bi{proactively disclose} the use of AI tools in content they share. Third, other bills state that \emph{no party} may \bi{remove disclosures} from AI content. Lastly, a few bills mandate or encourage that specific \emph{entities} must \bi{develop watermarking standards and guidelines}.

The values for this parameter are not mutually exclusive, as a bill may mandate multiple provisions simultaneously. 
\iffullversion
    In fact, as our examples later on will show, some of these are \emph{complementary}; for example, a bill may mandate watermarks on model outputs that follow to-be-determined standards. 
\fi
Further, not all legislative documents include mandates related to disclosures, as four documents only include mandates for the development of standards. For these, the values of parameters \param{4} through \param{6}, described below, are empty.

\begin{parameter}\label{param:technologies}
What disclosure mechanisms are in scope of the mandates?
\end{parameter}
The fourth parameter defines the \emph{technologies} or \emph{mechanisms} that comply with the disclosure-related mandates discussed under \param{3}. 
For example, a bill that states that model providers must add disclosures to outputs may specify that this disclosure must be of a certain form or type. 
We observed two categories within this parameter. Some bills are \bi{broad}, providing a high-level (and sometimes vague) specification of acceptable mechanisms (or no description at all).  In such cases, many mechanisms could fall under the umbrella of the bill; at the same time, it may not be clear what constitutes an adequate solution. 
The second category mentions \bi{specific} types of mechanisms (e.g., imperceptible watermarks).\footnote{We are pleased to report that no bills refer to specific technologies rather than \emph{types} of technologies (except in the form of examples). Writing specific technologies into law is generally inadvisable as it is brittle to future technological developments (e.g., \cite{nyt-sk}).}

\begin{parameter}\label{param:technologies-properties}
What properties must disclosure mechanisms satisfy?
\end{parameter}
The next parameter defines any \emph{requirements} that disclosure technologies (as scoped in \param{4}) must meet.
This parameter encompasses three categories. First, many bills specified \bi{no requirements}. Second, other bills had broad, flexible \bi{qualitative requirements}.
For example, such a constraint may state that disclosures on model outputs should be ``permanent or extraordinarily difficult to remove, to the extent it is technically feasible'' (e.g, \uidleg{S 942} in California). Lastly, some bills had \bi{specific requirements} in more precise technical terms, which define concrete constraints with little to no room for interpretation. 
\iffullversion
    For example, a bill of this type may say that disclosures added by users must be of a certain font size, and present in a specific location within the text, such as \uidleg{H 1734} in Hawaii.
\else
    For example, a bill of this type may say disclosures must have a certain font size, such as \uidleg{H 1734} in Hawaii.
\fi

\begin{parameter}\label{param:penalties-remedies}
What are the consequences for violation? 
\end{parameter}
Lastly, the sixth parameter relates to the \emph{consequences} for violating a bill's disclosure mandates. 
Bills instantiated this parameter in three ways. First, some bills specified \bi{monetary penalties}, such as a fee that must be paid for each individual violation (sometimes fixed, and other times increasing with the number of violations). Second, other bills defined penalties equivalent to \bi{those in other existing statutes}. Lastly, a few bills did not provide a specification of consequences.

\medskip
\iffullversion
    In summary, these six parameters collectively capture the key structural elements of a content transparency bill: the content that it covers (which type of AI-generated content and under which use cases) (\param{1}, \param{2});
    the way in which it regulates said covered content (which party must act in what way) (\param{3});
    the acceptable technical solutions (which technologies and under which requirements) (\param{4}, \param{5});
    and the penalties for violating it (\param{6}).
    
    In other words, our taxonomy asks ``who'' (\param{3}) must do something ``to what'' (\param{1}), in ``what way'' (\param{3}, \param{4}, \param{5}), and ``when'' (\param{2}); and ``what happens'' if not (\param{6}).
\else
    In summary, our framework asks ``who'' (\param{3}) must do something ``to what'' (\param{1}), in ``what way'' (\param{3}, \param{4}, \param{5}), and ``when'' (\param{2}); and ``what happens'' if not (\param{6}).

    We apply our framework to each legislative document in our corpus, and extract a comparison-friendly systematization capturing the key elements of existing legislative efforts. Due to space constraints, the full systematization of all 43 bills is available at~\cite{systematizationRepo}. 
    We present two detailed examples in Appendix~\ref{a:bill-framework-examples}, which are also summarized in Table~\ref{tbl:bill-framework-examples}.
    We present two examples in Table~\ref{tbl:bill-framework-examples}, the first of which is explained in detail in Appendix~\ref{a:bill-framework-examples}. (We omit a detailed debrief of the second one due to space constraints.)
    \fi

\iffullversion
    \subsection{Applying Our Taxonomy}\label{subsec:archetypes}

    We apply our taxonomy to each legislative document in our corpus, and thus extract a comparison-friendly systematization capturing the key elements of existing legislative efforts related to watermarking. The full systematization of all 43 bills is given in an external repository.\footnote{\url{https://anonymous.4open.science/r/watermarking-policy-C23E/README.md}}
    We present two illustrated examples, which are also summarized more briefly in Table~\ref{tbl:bill-framework-examples}.
     \newcolumntype{B}{X}
\newcolumntype{S}{>{\hsize=.5\hsize}X}

\begin{table*}[t!]
    \centering
    \footnotesize
    \renewcommand{\arraystretch}{1.1}
    \resizebox{\linewidth}{!}{%
    \begin{tabular}{l|llllll}
     \toprule
     \textbf{Document} &\textbf{\param{1}. Content scope} & \textbf{\param{2}. Use cases} & \textbf{\param{3}. Mandates} & \textbf{\param{4}. Mechanisms} & \textbf{\param{5}. Requirements} & \textbf{\param{6.} Penalties} \\
     \midrule
     \makecell[l]{\uid{S 942} \\ (California)} & \makecell[l]{Unqualified usage \\ of AI tools} & \makecell[l]{General (with \\ exceptions)} & \makecell[l]{Providers must add disclosures \\ and provide a detection tool} & Broad & \makecell[l]{Qualitative \\ and specific} & \makecell[l]{Monetary \\ penalties} \\
     \midrule
     \makecell[l]{\uid{S 4674} \\ (Congress)} & \makecell[l]{Qualified usage of AI \\ tools (quantitative)} & General & \makecell[l]{Providers must add disclosures, \\ disclosures must not be removed, \\ and standards must be developed} & Broad & Qualitative & \makecell[l]{Violation \\ of statute} \\
     \bottomrule
\end{tabular}}
\caption{\textbf{Example summaries of two legislative documents}, by parameter values. See \secref{subsec:archetypes} for a more detailed description of these examples. The full systematization of all legislative documents can be found in our external repository~\cite{systematizationRepo}.}
\label{tbl:bill-framework-examples}
\end{table*}

    \paragraph{Example \#1: S 942 (California).} This bill from the California Senate defines a number of disclosure requirements for providers of GenAI models. Notably, it is one of the 7 bills in our list that has been enacted into law. 
    \begin{newitemize}
        \item \emph{\param{1} (content scope):} Covered content is defined by the \emph{unqualified} usage of AI. Concretely, AI is defined as ``an engineered or machine-based system that varies in its level of autonomy and that can, for explicit or implicit objectives, infer from the input it receives how to generate outputs that can influence physical or virtual environments''. Then, a ``GenAI system'' is defined as ``artificial intelligence that can generate derived synthetic content, including text, images, video, and audio, that emulates the structure and characteristics of the system's training data''. Lastly, covered content is defined as that which is ``created or altered'' by a GenAI system.
        \item \emph{\param{2} (use cases):} The bill has \emph{general} applicability, but \emph{excludes} ``any product, service, internet website, or application that provides exclusively non-user-generated video game, television, streaming, movie, or interactive experiences''.
        \item \emph{\param{3} (mandates):} The bill mandates that \emph{model providers} (as per defined characteristics) must \emph{include disclosures} on outputs of their models, and provide publicly accessible detection tools.
        \item \emph{\param{4} (technologies):} Detection technologies are \emph{broad}, simply mentioning that disclosures must be ``latent disclosures'' and (if the user opts) ``manifest disclosures''. For example, imperceptible and metadata watermarks could satisfy the former, and perceptible watermarks the latter.
        \item \emph{\param{5} (technology requirements):} The bill includes a mix of both \emph{qualitative} and \emph{specific} requirements. For the former, disclosures should be ``permanent or extraordinarily difficult to remove, to the extent it is technically feasible''. For the latter, the bill specifies a concrete list of metadata that must be included in the disclosure, as well as the fact that the detection tool must be accessible through both a URL and API.
        \item \emph{\param{6} (penalties):} \emph{Monetary penalties} of \$5,000 for each violation.
    \end{newitemize} 
    
    \paragraph{Example \#2: COPIED Act (Congress).}  This federal  bill from the U.S. Senate introduces a number of transparency requirements on AI content. Unlike \uidleg{S 942}, this bill was not enacted into law.
    
    \begin{newitemize}
        \item \emph{\param{1} (content scope):} The definition of AI is referenced from 15 U.S.C. \S\ 9401, which defines AI as ``a machine-based system that can, for a given set of human-defined objectives, make predictions, recommendations or decisions influencing real or virtual environments''. Then, covered content is defined by the \emph{qualified} usage of AI, specified \emph{quantitatively}: content which is ``wholly generated'' or ``significantly modified'' by ``algorithms, including artificial intelligence''.
        \item \emph{\param{2} (use cases):} The bill has \emph{general} applicability.
        \item \emph{\param{3} (mandates):} The bill mandates that \emph{model providers} must \emph{include disclosures}, disclosures must \emph{not be removed}, and \emph{standards} must be developed.
        \item \emph{\param{4} (technologies):} Technologies are \emph{broad}, simply mentioning ``content provenance information''.
        \item \emph{\param{5} (technology requirements):} The provenance information should be ``to the extent technically feasible, reasonable security measures to ensure that such content provenance information is machine-readable and not easily removed, altered, or separated from the underlying content'', which corresponds to a \emph{qualitative} requirement.
        \item \emph{\param{6} (penalties):} Violations ``shall be treated as a violation of a rule defining an unfair or deceptive act or practice prescribed under section 18(a)(1)(B) of the Federal Trade Commission Act (15 U.S.C. 57a(a)(1)(B))''.
    \end{newitemize}
    
    A repository containing our full systematization of legislative documents, in the style of the two examples above, can be found online.\footnote{\url{https://anonymous.4open.science/r/watermarking-policy-C23E/README.md}}
    
\fi

\iffullversion
    \iffullversion
    \subsection{Types of Legislative Patterns}
    \label{subsec:patterns}
\else
    \section{Types of Legislative Patterns}
    \label{a:patterns}
\fi

Our systematization revealed \emph{three main patterns} in legislative documents\iffullversion\else (briefly highlighted in Section ~\ref{subsec:patterns})\fi: calls for (1) explicit disclosures from entities who use AI tools in particular use cases, (2) disclosures from model providers for all outputs, and (3) developing standards.  

We explain these in this section at a high level, and expand on them as part of our analysis in Sections~\ref{subsec:leg-takeaways}. These patterns emerge from our systematization in the form of three frequently-repeating parameter configurations, 
as we summarize in Table~\ref{tbl:patterns-summary}.

\newcolumntype{B}{X}
\newcolumntype{S}{>{\hsize=.5\hsize}X}

\begin{table}
    \centering
    \iflaw\else\footnotesize\fi
    \renewcommand{\arraystretch}{1.1}
    \begin{tabular}{l|llll}
     \toprule
     \textbf{Pattern} &\textbf{\param{1}} & \textbf{\param{2}} & \textbf{\param{3}} & \textbf{\param{4}} \\
     \midrule
     \pattern{1} & N/A & Specific & \makecell[l]{Users must \\ add disclosures} & \makecell[l]{Explicit \\ disclosures} \\
     \midrule
      \pattern{2} & N/A & General & \makecell[l]{Providers must \\ add disclosures} & N/A \\
     \midrule
     \pattern{3} & N/A & N/A & \makecell[l]{Government agencies \\ must develop content \\ transparency standards} & N/A \\
     \bottomrule
\end{tabular}
\caption{Overview of how the three patterns of \secref{subsec:patterns} map to the parameters of our framework from \secref{subsec:bills-framework}. All three patterns map to ``N/A'' for parameters \textbf{\param{5}} and \textbf{\param{6}} so we omit these from the table.}
\label{tbl:patterns-summary}
\end{table}

\paragraph{Legislative pattern \#1 (\pattern{1}): user disclosures.} A common pattern in watermarking bills is to \emph{mandate that entities who use AI tools to create or modify content for a specific use case must add a conspicuous disclaimer to such content}. This corresponds to instantiations where \param{2} is ``specific,'' \param{3} is ``model users must add disclosures,'' and \param{4} is some type of explicit disclosure, such as labels on content. For example, \uidleg{S 88} from North Carolina, says that political advertisements created ``in whole or in part'' using AI need to include an explicit statement disclosing its use. What exactly constitutes AI content (i.e., \param{1}) varies across bills, generally revolving around the degree to which AI tools were used; we analyze this in the next subsection. Bills of this type also generally  impose additional requirements on the disclosure, such as, in the case of text, constraints on its color or font size (e.g., \uidleg{S 1044} in Pennsylvania).

\paragraph{Legislative pattern \#2 (\pattern{2}): provider disclosures.} The next common pattern we noticed is to \emph{mandate that model providers must embed a disclosures in all outputs generated by their models}. This corresponds to instantiations where \param{2} is ``general'' and \param{3} is ``model providers must add disclosures''. The specific  mandated disclosure mechanism (\param{4}) and any requirements on it (\param{5}) vary across such bills. An example of a bill that follows this pattern is \uidleg{S 217} in Ohio, which says that model providers must include a ``distinctive watermark'' on any content produced by their models. In some (but not all) cases, bills that follow this pattern additionally include ``criminalize removal of disclosures'' in  \param{3}; no bills that exclusively follow patterns \#1 or \#3 include this mandate. 

\paragraph{Legislative pattern \#3 (\pattern{3}): standards.} Lastly, the third common pattern is to \emph{mandate the development of standards}. This corresponds to an instantiation of \param{3} as ``government agencies must develop content transparency standards''.

Of the 43 legislative documents, twenty-five followed \pattern{1} only, eight followed \pattern{2} only, and four followed \pattern{3} only. Five followed \pattern{2} and \pattern{3}, and one followed all thee patterns. Table~\ref{fig:legislative-docs-summary} gives a complete overview of which bills followed which patterns.

\else
    \subsection{Types of Provisions}
    \label{subsec:patterns}
    
    Our systematization revealed \emph{three main types} of legislative patterns that characterize the content of legislative documents. These patterns emerge from our systematization in the form of three frequently-repeating parameter configurations, pointing to the benefits of our structured approach.
    \begin{newitemize}\label{itemize:patterns-summary}
        \item \textbf{Legislative pattern \#1 (user disclosures):} entities who \emph{use AI tools} to create or modify content for a \emph{specific use case} must add a \emph{conspicuous disclaimer} to such content.
        \item \textbf{Legislative pattern \#2 (provider disclosures):} \emph{model providers} must \emph{embed a disclosure} in all outputs. %
        \item \textbf{Legislative pattern \#3 (standards):} government agencies should \emph{develop standards}.
    \end{newitemize}

    Due to space, further elaboration of these patterns (and how they map to our framework's parameters) is in Appendix~\ref{a:patterns}. 
    We show the distribution of legislative patterns across bills in Table~\ref{tbl:legislative-docs-summary}. Moving forward, we refer to them as \pattern{1}, \pattern{2}, and \pattern{3}, respectively.
\fi

\subsection{Key Trends and Open Questions}\label{subsec:leg-takeaways}
Next, we distill the key trends and themes in our legislative documents, with special attention to gaps between policy and practice: points of ambiguity, potential pitfalls, and the disconnect between legislative language and technological limitations. Along the way, we identify open questions that are critical for effective watermarking policy.

\paragraph{Ambiguities in covered content.} We begin by analyzing trends in the ways in which legislative documents define their scope of covered content (\param{1}). 

\begin{takeaway}\label{kt:scope-room}
    Legislative documents tend to leave considerable room for interpretation in their scope of covered content. 
\end{takeaway}

A common instantiation of \param{1}, appearing in seventeen bills, is ``unqualified usage of AI tools''; that is, content is in scope of these bills if it has been modified or generated by AI in any way. 
\iffullversion
        Bills of this type typically contain expressions such as ``generated in whole or in part with the use of artificial intelligence'' (e.g., \uidleg{A 7904} in New York and \uidleg{H 1267} in Mississippi), or simply ``any output generated by artificial intelligence'' (e.g., \uidleg{HR 3831} in Congress). 
    
\fi
AI tools can be and are routinely used to modify content in a myriad of common ways, sometimes unbeknownst to users. Targeting \emph{all} these types of uses may not be desirable or intended 
\iffullversion
    even in bills defining their scope of covered content broadly. 
\else
    in all these bills.
\fi
For example, modern spelling and grammar corrections are often AI-based, and modern cellphone cameras fairly ubiquitously apply AI-based enhancement to photos (e.g., adjusting brightness).
Nuanced consideration of varied uses of AI, especially where AI is embedded in other applications and thus less visible than in standalone AI applications or applications advertised as ``AI-powered,'' will be essential for judicial interpretation of such broad legislative language.

The other bills have more specific scopes, corresponding to instantiations of \param{1} as ``qualified usage of AI tools''. Such bills explicitly exclude certain kinds of alterations from their scope. 
\iffullversion
    Qualifications on AI use are specified in two main ways. 
First, eight bills narrow their scope to content that has been \emph{substantially modified} by AI. 
For example, \uidleg{A 664} in Wisconsin specifies \param{1} as content ``substantially produced in whole or in part'' using AI. While ``substantially'' is not always defined in bills of this type, this already provides some high-level guidance for their interpretation. Fifteen bills go a step further, narrowing their scope with concrete definitions for what constitute substantial modifications. First, five bills defined substantial modifications as those that change the \emph{perceived meaning} of content. For example, \uidleg{S 2691} by Congress defines its scope as usage of AI that ``materially alters the meaning or significance that a reasonable person would take away from the content''. 
Another ten bills define substantial modifications as those that change the \emph{truth provenance} of content, i.e., using AI to produce fake content that appears real. For example, \uidleg{A 2355} in California specifies \param{1} as AI generated content that ``would falsely appear to a reasonable person to be authentic''. 

\else
    \phantomsection\label{reftoappendix:substantial-modification}
    A common approach (adopted by 23 bills) is to narrow scope to content that has been \emph{substantially modified} by AI. Appendix~\ref{a:substantial-modification} discusses how substantial modification is defined, which varies across these bills.
\fi

\begin{takeaway}\label{kt:scope-modified}
    Bills that narrow their scope to content that is ``substantially modified'' by AI often do so in terms of changes to the content's perceived meaning or truth provenance.
\end{takeaway}

There is a notable difference between defining substantial modifications in terms of perceived meaning vs truth provenance: for the latter, \emph{inauthentic-looking content need not require a disclosure}. Therefore, content that is clearly false but has real-looking components, and thus should probably be marked as AI-generated, is a potential gray area. Moreover, content that is clearly inauthentic, which can still cause serious harms (e.g., pornographic deepfakes~\cite{sobel2024real}), is out of scope. That said, of course, a bill may focus on a particular area of harms related to AI-generated content, and need not be expected to address all of them.

\begin{openQuestion}\label{openq:content-scope}
    How should we determine what constitutes `AI-generated' content? To what extent should such determinations be context-dependent?
\end{openQuestion}

\paragraph{Burden of disclosure responsibility.} 
\iffullversion
    A fundamental question for legislative documents is \emph{who} will be responsible for disclosing content as AI-generated: model users or model providers (\param{3}).
    In our legislative corpus, the most common approach, appearing in 26 documents,\footnote{Including 5 out of 7 enacted laws.} is that model \emph{users} 
    (rather than providers)
    must include disclosures on AI-generated content.
\fi

\begin{takeaway}\label{kt:onus-disclosure-users}
    A majority of legislative documents place the onus of disclosure on the users of AI tools.
\end{takeaway}

\iffullversion
    All bills that place disclosure responsibility on users include a specific context of applicability (\param{2}), corresponding to \pattern{1} from \secref{subsec:patterns}. 
\else
    All bills that place disclosure responsibility on users include a specific context of applicability (see \param{2} and \pattern{1}).
\fi
By far the most common one is \emph{election and political communications and advertising}, appearing in 17 bills, suggesting heightened political concern in this context.

Relying on user disclosures leads to several challenges. It significantly increases the number of entities covered by the mandates, raising questions around how violations will be tracked and how the law will be enforced. Additionally, it raises complex questions regarding individual human rights, such as anonymity online and freedom of expression.

Fourteen bills instead mandate that model \emph{providers} must add disclosures to content generated by their models. 
\iffullversion
    These bills do not specify a specific context of applicability, which corresponds to \pattern{2} from \secref{subsec:patterns}.
    Broad mandates of disclosures from model providers requires a nuanced consideration of the practical limitations of technologies and AI models.
\else
    These bills do not specify a specific context of applicability (see \param{2} and \pattern{2}).
    Broad mandates of disclosures from model providers requires a nuanced consideration of the practical limitations of technologies and AI models.
\fi 

\begin{takeaway}\label{kt:onus-disclosure-providers}
    In contrast to computer-science literature, which generally focuses on properties of \emph{models}, bills often impose blanket requirements on \emph{model outputs}. This discrepancy can make proposed requirements ambiguous or technically unachievable when literally interpreted.

\end{takeaway}

For example, in many common scenarios, it is impossible for watermarking schemes to tag all outputs. Intuitively, in situations where there is little flexibility for variation in the model's responses,\footnote{For example, ``output a single vowel'' (there are only 5 options) or ``what is the opening sentence of Moby Dick?'' (there is only one correct answer).} the lack of flexibility may mean it is technically impossible to add a watermark, especially if the watermark must contain significant amounts of information.\footnote{For example, it is not possible to embed a date and time into a single character: there is just not enough room. The output text must be at least as long as the embedded information in the watermark. Of course, watermark information could be added at the end of such an output --- but then, it would be straightforward to remove, and therefore fail to satisfy robustness.} Technical definitions of watermarking therefore define requirements on models along the lines of ``embed a watermark whenever conditions X and Y are satisfied'', recognizing the impossibility of embedding a watermark in every output. 

We found that every bill that placed the onus of disclosure on model providers stated broad requirements on model outputs, creating ambiguity\iffullversion, as a literal interpretation of such requirements could be technically impossible to satisfy.\else.\fi

Mandating provider disclosures also raises important questions regarding public release and use of models. In the case of ``open'' models (e.g., Meta's Llama~\cite{touvron2023llama}), a model's creator may be different from the hosts of a (possibly modified) version of the model.  %
\iffullversion
    In such cases, it may be unclear who is responsible for unlabelled outputs.
    Furthermore, for both open and closed models, it is common for users to interact with them through intermediary applications, who then communicate with models in the backend. Once again, it is not clear whether and to what extent intermediary application providers can be held responsible as `providers', and how disclosure responsibilities should be shared between different providers within a software ecosystem. 
    (We discuss open models further in \secref{sec:reg}.)
\else
    Moreover, users commonly interact with models through intermediary applications.
    In such cases, it may be unclear who is responsible for unlabelled outputs: the creator, the host, the intermediary, or a combination thereof?
\fi

\begin{openQuestion}
\label{openq:responsibility}
    Who should bear the responsibility of disclosure in which contexts, and how should responsibility be divided between parties?
\end{openQuestion}
\iffullversion
    Whether users or providers are the ones responsible, it is important that mandates are achievable from a technical standpoint, being compatible with the limitations and practical considerations of the available solutions, and that they take into account the ways in which AI models are used and distributed in practice. %
\fi

An important related question concerns who should bear the responsibility for secure and correct design, implementation, and deployment of content transparency infrastructure. Notably (and uniquely among our Open Questions), this question does not appear in our corpus. \secref{sec:recommendations} provides additional discussion.

\begin{openQuestion}
\label{openq:trust}
    Who are we relying on for content transparency infrastructure?
\end{openQuestion}

\paragraph{Criminalizing removal of disclosure information.} 
Seven bills mandate that no party may \emph{remove} disclosures. 
\iffullversion
    (As noted in \secref{subsec:patterns}, all such bills correspond to \pattern{2} from \secref{subsec:patterns}.)
\fi
Criminalizing removal of disclosures raises a number of serious risks. Existing and foreseeable watermarking technologies are fairly easily removable,\footnote{That is, they do not and practically cannot have perfect \emph{robustness}.} whether accidentally or on purpose\cite{zhang2024watermarks}. For example, watermarks can often be removed by commonplace activity like automatic whitespace normalization, image compression, etc.--- which could be criminalized under these bills. Also, research to thoroughly understand and improve the robustness of watermarks necessarily involves studying watermark removal techniques and may involve watermark removal.
We stress the importance of avoiding criminalizing such research, which is essential to our understanding of watermarking technologies.

\begin{takeaway}\label{kt:removal}
    Many bills prohibit the removal of disclosures, which could criminalize commonplace activities and essential content transparency research.
\end{takeaway}

\iffullversion
    From the seven bills, only \uidleg{S 217} in Ohio had additional qualifications, stating that removals must be ``with the purpose of concealing'' that content is AI generated. Even then, it is difficult to draw a line between accidental and adversarial removal of disclosure information, and intentional experimentation with removal techniques is essential to certain research activities. (In \secref{sec:recommendations}, we describe other possible issues with criminalization of removal of disclosure information.)
\fi

\begin{openQuestion}
\label{openq:criminalizing}
    Can criminalization of disclosure removal avoid criminalizing commonplace activities and chilling essential content transparency research?
\end{openQuestion}

\iffullversion
    Imperfect robustness in disclosure technologies is unavoidable~\cite{zhang2024watermarks}, and watermarking legislation must consider this imperfection in determining when, where and how to prohibit removal of disclosures if at all.
\fi

\paragraph{Disclosure requirements are not clear cut.} Required types of disclosures (\param{4}) vary significantly. 

A prominent trend is that bills often leave notable ambiguity in what forms of disclosure would suffice.
\iffullversion
    All documents that mandate user disclosures require some form of \emph{perceivable} disclosure in \param{4}, appropriate for the type of content (e.g., visible labels for text, a voice recording for audio, etc).\footnote{An interesting instance of pattern \#1 is \uidleg{H 1147} in Colorado, one of the seven bills enacted into law. Unlike all other bills of this type, in addition to a perceivable disclosure, it requires that model users embed provenance information in the \emph{metadata} of content, including ``the identity of the tool'' used to create the content, and the date and time of creation. This requires a high degree of technical proficiency, likely not possessed by most model users.} 

    However, detailed disclosure requirements are rarely specified precisely, with most bills using high-level terms such as ``clear and conspicuous'' disclosures (\uidleg{S 1044} in Pennsylvania), ``clear and prominent disclaimer'' (\uidleg{H 1267} in Mississippi), and so on.
    These correspond to instantiations of \param{4} as ``broad'', and so many technologies may satisfy these mandates, such as labels next to content, overlaid watermarks, and more. %

    The bills that mandate model provider disclosures (\pattern{2} in \secref{subsec:patterns}) similarly tend to endorse broad categories of disclosure while eliding specifics. For example, four bills\footnote{\uidleg{S 2} from CT, \uidleg{S 4674} and \uidleg{HR 7766} from Congress, and the \uidleg{EU AI Act}.} define their required technologies as ``machine-readable'' disclosures. Of these, only \uidleg{HR 7766} includes a definition of ``machine-readable,'' citing 44 U.S.C. \S 3502.\footnote{``[I]n a format that can be easily processed by a computer without human intervention while ensuring no semantic meaning is lost.''} Another example is \uidleg{S 942} in California, enacted into law, which requires ``manifest'' and ``latent'' disclosures, defined respectively as ``easily perceived, understood, or recognized by a natural person'' and ``present but not manifest.'' %

    The least specific bill is \uidleg{HR 3831}, which simply states that model outputs must include the text 
    \iflaw\else\begin{quote}\fi
    ``Disclaimer: this output has been generated by artificial intelligence''    
    \iflaw\else\end{quote}\fi
    with no additional details of where and how this disclaimer should be embedded. %
\else
    All documents that mandate user disclosures require some form of \emph{perceivable} disclosure in \param{4} (see \pattern{1}).\footnote{E.g., visible labels for text, a voice recording for audio, etc.}
    However, detailed disclosure requirements are rarely specified precisely, with most bills using high-level terms such as ``clear and conspicuous'' (\uidleg{S 1044}), ``clear and prominent disclaimer'' (\uidleg{H 1267}), and so on.
    The bills that mandate model provider disclosures similarly tend to elide
    specifics.\footnote{The least specific bill is \uidleg{HR 3831}, which simply states that model outputs must include the text ``Disclaimer: this output has been generated by artificial intelligence'', with no additional details on how, where, etc.}
\fi

\phantomsection\label{reftoappendix:watermark-def}
Another notable trend is that various bills define their required technologies with references to the often ambiguous term ``watermark''. Five bills use the term with no definition or explanation of what it entails. 
Two other bills, H 710 in Vermont and H 2660 in Pennsylvania, do provide definitions. The former defines watermarking as ``information that is embedded in, and reasonably difficult to remove from, any digital content; and enables a consumer [...] to verify the authenticity of the digital content and to determine whether the digital content is synthetic digital content''. The latter defines it as ``a mark placed on an artificial intelligence generated image, simulation or video''.\footnote{Confusingly, while this definition only mentions images and video, AI-generated text is also in the content covered by this bill.} Further reading of 2660 suggests implicitly that it may require visible watermarks, since there are a number of requirements (\param{5}) related to how it should be perceived, such as covering 30\% of the content with minimum of 50\% opacity. Lastly, two bills, the EU AI Act and AB 3211 in California, refer to ``watermarks'' as examples of broader permitted classes of disclosures: ``marking in a machine readable format'' for the former; and ways to communicate ``provenance data'' which ``records the origin or history of digital content'' for the latter.

\begin{takeaway}\label{kt:watermark-ambigous}
    ``Watermark'' is often used ambiguously within and across bills. Disclosure requirements vary considerably, and often are not clear cut.    
\end{takeaway}

The space of content transparency mechanisms is vast and complex. Different technologies lead to different guarantees, tradeoffs, and practical considerations. Ideally, identifying which technologies suffice to meet a bill's requirements should be straightforward; and the way in which bills should state these requirements is therefore critical.

\begin{openQuestion}\label{openq:disclosure-types}
    Which content transparency mechanisms are best suited to achieve the goals of disclosure in different contexts? How should legislative documents specify these clearly while maintaining flexibility to adapt to future technological change?
\end{openQuestion}

\iffullversion
    The initial technical question is to understand the range of possible disclosure technologies and their limitations, and evaluate their fit for a given legislative context. Then, a distinct and important challenge lies in articulating these requirements in legislation with clarity. Legislation needs to navigate the delicate balance between maintaining flexibility and ``technology neutrality''~\cite{greenberg2015rethinking,reed2007taking}, while avoiding vagueness, which leads to uncertainty for model providers and end users. 
\fi

\paragraph{Incompatible technological requirements.} 
\iffullversion
    Some documents include additional requirements  on disclosures (\param{5}), such as specifying clearer requirements for formats and types of disclosure. But as we explain below, these requirements are sometimes so broadly stated that, if literally interpreted, they are not compatible with the practical limitations of many current content transparency technologies.
    
    Thirty-five bills impose at least one requirement on their mandated solutions.\footnote{From the other seven bills, three follow only pattern \#3 (development of standards) and four follow only pattern \# 1 (user disclosures).} We organize these requirements into four categories: \emph{cosmetic}, \emph{informational}, \emph{operational}, and \emph{performance}, and explain each below.
\else
    Some documents include additional requirements  on disclosures (\param{5}), such as specifying formats and types of disclosure.
    We organize these into 4 categories: \emph{cosmetic}, \emph{informational}, \emph{operational}, and \emph{performance}.
    Such additional requirements are sometimes so broadly stated that, if literally interpreted, they are incompatible with the practical limitations of many current content transparency technologies.
\fi

\iffullversion
    \emph{Cosmetic} requirements specify constraints on the way disclosures must be displayed. For example, for perceptible disclosures on text, this may consist of a certain font size or a certain location with respect to the text. By definition, cosmetic requirements are only relevant for bills that mandate some type of perceivable disclosure in \param{4}. From the twenty-eight such bills---all twenty-five following \pattern{1}, and three following \pattern{2}---24 have at least one cosmetic requirement.
\else
    \emph{Cosmetic} requirements specify constraints on the way disclosures must be displayed (e.g., font size or location).
\fi

\iffullversion
    \emph{Informational} requirements specify information that must be included in the disclosure.
    29 bills  have at least one informational requirement. There are three main types of informational requirements. The most basic type, found in 21 bills, is a generic statement disclosing that the content is AI-generated, such as ``This image has been manipulated or generated by artificial intelligence'' (\uidleg{S 880} in North Carolina). Second, 8 bills additionally require \emph{model information} as part of the disclosure; for example, \uidleg{H 5321} in Illinois requires a specification of ``the identity of the tool''. Lastly, four bills further require \emph{creation metadata} in the disclosure, such as the date and time of content creation. Three such bills---\uidleg{S 942} in California, \uidleg{H 1147} in Colorado, and \uidleg{S 131} in Utah---are among the six bills enacted into law; the first two require date and time of content creation, and the latter ``the initial author and creator of the content'' as well as ``subsequent entities'' that modified the content.
\else
    \emph{Informational} requirements specify information that must be included in disclosures, and come in 3 main types: statements describing content as AI-generated (e.g, \uidleg{S 880}), model information (e.g., \uidleg{H 5321}), and creation metadata~(e.g.,~\uidleg{S 942}).
\fi

Informational requirements raise two important challenges. First, requiring creation metadata may be in tension with user privacy, since information like the creator and date/time/location of creation can be sensitive. 
Second, the amount of information to be disclosed may exceed the capabilities of certain disclosure technologies, which have limits on the amount of information they can include, such as imperceptible watermarks, which have a \emph{bit rate} that defines the maximum amount of information that can be included~\cite{zhao2025watermarking}. %

\iffullversion
    As such, requiring a lot of information in disclosures may exclude many natural solutions, leading to tensions between mandates and requirements. For example, \uidleg{S 942} in California requires ``latent disclosures''\footnote{Defined as ``present but not manifest'', where ``manifest'' is defined as ``easily perceived, understood, or recognized by a natural person''.} from model owners, as well as all three types of informational requirements. In practice, imperceptible watermarks are one prominent type of latent disclosure. However, imperceptible watermarks' bit rate varies depending on the medium, which may not be sufficient to fit all the required information: for text, for example, schemes can typically only include a few bits of information~\cite{zhao2025watermarking}.
\fi

\begin{takeaway}\label{kt:information}
    Many bills require that specific information is included in disclosures, such as model information or creation metadata. This may reveal sensitive information, or constrain the technologies that may be used to satisfy the bill's requirements.
\end{takeaway}

\iffullversion
    The third category, \emph{operational} requirements, specify details about the functioning of disclosure mechanisms. Such requirements are most relevant for ``invisible'' disclosures (e.g., imperceptible watermarks, metadata tagging, etc), which are embedded latently in content and require specific \emph{detection processes} to verify their presence, such as specialized algorithms. Detection methods involve a number of nuances and considerations with important policy implications: What specific information do detection tools reveal? Can detection be performed by anyone, or only designated people (if so, who)? In what ways must detection tools be accessible? Should they be implemented by the model provider, or a third party? And so on. But only two out of the fourteen bills that require some form of imperceptible disclosure acknowledge a detection process at all. The other bills would leave open key questions around detection mechanisms and therefore around the utility of the watermarks. A literal interpretation of the bills omitting mention of detection could mean, in principle, that an imperceptible watermark that is not detectable (and is therefore useless) would meet the mandates of the bill. 

    The two bills that do mention a detection process---\uidleg{AB 3211} and \uidleg{S 942} in California---do so at a high level of generality, stating that the tool must be compatible with the required disclosures, and must output any provenance information in the content. Both of these bills also require date and time of creation in the disclosures, which means that said detection tools would reveal this information. \uidleg{S 942} adds a few more requirements to detection tools, stating that it must be accessible by the provider's website or an API.
\else
    The third category, \emph{operational} requirements, specify details about the functioning of disclosure mechanisms. Such requirements are most relevant for ``invisible'' disclosures (e.g., imperceptible watermarks, metadata tagging, etc), which are embedded latently in content and require specific \emph{detection processes} to verify their presence. 
    Detection methods involve a number of nuances and considerations with important policy implications: What information do detection tools reveal? Who can perform detection? Who implements detection tools? But only two of the 14 bills that require imperceptible disclosure acknowledge a detection process at all. The other bills necessarily leave open key questions around detection mechanisms and therefore around the utility of the watermarks.
\fi

\begin{takeaway}\label{kt:detection-process}
    Most bills do not acknowledge detection processes for imperceptible disclosure mechanisms.
\end{takeaway}

\phantomsection\label{reftoappendix:robustness}
Lastly, \emph{performance} requirements specify constraints on the various properties of technologies, such as quality, false positive/negative rate, robustness, unforgeability, etc~\cite{zhao2025watermarking}. Some of these properties have inherent limitations or are in tension with others, offering a constrained space of tradeoffs. Across all documents, however, the only technical property acknowledged (explicitly or implicitly) is \emph{robustness}, 
\iffullversion
    mentioned in 10 documents. 

    For the documents that do acknowledge robustness, some interesting trends emerged. First, all documents reference robustness in a broad, general manner, such as requiring disclosures that are ``permanent or unable to be easily removed'' (\uidleg{H 5321} in Illinois). Yet, robustness must typically be calibrated \emph{with respect to certain types of modifications}, leading to security tradeoffs against different types of edits. Mandating robustness in a general sense thus leaves developers with no guidance on how to choose between multiple alternative approaches whose guarantees differ very substantially. 

Six bills add more qualifications to their robustness requirements, mandating that disclosures are as robust ``as technically feasible'' (e.g., \uidleg{S 2691} by Congress).
This type of requirement raises two key challenges.
First, it sets a moving target, with expectations rapidly changing as the state-of-the-art develops. 
Furthermore, there are many notions of robustness, whose levels of strength are incomparable\footnote{For example, an image watermark may be robust to cropping but easily removed by compression; this is not clearly stronger or weaker than a watermark robust to compression but removed by cropping.}.
Second,
robustness is in inherent tension with other desirable properties (such as output quality) so disclosures that are maximally robust will significantly degrade other properties; for example, a visible watermark that is maximally robust would be one that completely obfuscates the original content --- that is, renders the original content useless.
While it seems unlikely that ``optimize for robustness at the cost of everything else'' is what these bills intend, and unlikely that courts would it interpret them thus, the bills' robustness standards remain unclear.

\else
    mentioned in 10 documents (see Appendix~\ref{a:robustness} for details).
\fi

\begin{takeaway}\label{kt:properties}
    Bills do not engage in depth with essential security properties of content transparency mechanisms (such as robustness) and the tradeoffs between them.
\end{takeaway}

\iffullversion
    It is important for legislative documents to acknowledge and harmonize with security properties and other practical considerations. How to make clear and actionable requirements while maintaining technology neutrality in this domain comes up again as a critical open question, as has also been an issue in many other areas of technology policy.\footnote{See, e.g., \cite{greenberg2015rethinking,reed2007taking}.} One possible approach, exemplified in some of the documents calling for standards, may be to acknowledge these properties at a high level but defer their specification to standardization agencies. 
\fi

\begin{openQuestion}\label{openq:reqs}
    What technical requirements are most relevant to content transparency policy, and how should legislative documents acknowledge them?
\end{openQuestion}

\paragraph{How to identify violations with certainty?} 
Currently, no methods exist to unambiguously identify content as AI-generated---motivating the need for content transparency legislation to begin with---which raises important concerns around any detection methods that are used to evidence a violation of a content transparency regulation. Enforcement must account for the 
(inherent) unreliability of detection.

\begin{takeaway}\label{kt:violations}
    It is unclear how violations can and will be determined, and how enforcement will account for the limitations of content transparency mechanisms. 
\end{takeaway}

\begin{openQuestion}
\label{openq:imperfect-detection}
    How can violations be determined reliably while taking into account the limitations of detection and labeling technologies?
\end{openQuestion}

\section{Analysis of Other Policy Documents}
\label{sec:reg}

\begin{table*}[ht!]
    \centering
    \resizebox{\linewidth}{!}{%
    \begin{tabular}{llrlllcl}
     \toprule
     \textbf{Identifier} & \textbf{Title} & \textbf{Year} & \textbf{Agency} & \textbf{Type} & \textbf{Revoked} & \textbf{Source}\\
     \midrule
    \uid{NIST AI 100-4} & \makecell[l]{Reducing Risks Posed \\ by Synthetic Content} & 2024 & \makecell[l]{National Institute of Standards \\ and Technology (NIST)} & Guidance & \noTbl & \makecell[l]{\href{https://doi.org/10.6028/NIST.AI.100-4}{URL} \\ (\href{https://perma.cc/MU7G-5USZ}{permalink})} \\
    \uid{EO 14110} & \makecell[l]{Executive Order on the Safe, Secure, \\ and Trustworthy Development and Use \\ of Artificial Intelligence} & 2023 & White House & Executive Order & \yesTbl & \makecell[l]{\href{https://www.federalregister.gov/documents/2023/11/01/2023-24283/safe-secure-and-trustworthy-development-and-use-of-artificial-intelligence}{URL} \\ (\href{https://perma.cc/U86V-H4XW}{permalink})} \\
    \uid{CISA 2024} & \makecell[l]{Risk in Focus: Generative A.I. \\ and the 2024 Election Cycle} & 2024 & \makecell[l]{Cybersecurity and Infrastructure \\ Security Agency (CISA)} & Briefing & \noTbl & \makecell[l]{\href{https://www.cisa.gov/sites/default/files/2024-01/Consolidated_Risk_in_Focus_Gen_AI_Elections_508c.pdf}{URL} \\ (\href{https://perma.cc/MY5B-JFSL}{permalink})} \\
    \uid{FCC 2024} & \makecell[l]{Disclosure and Transparency of Artificial \\ Intelligence-Generated Content \\ in Political Advertisements} & 2024 & \makecell[l]{Federal Communications \\ Commission (FCC)} & Article & \noTbl & \makecell[l]{\href{https://www.federalregister.gov/documents/2024/08/05/2024-16977/disclosure-and-transparency-of-artificial-intelligence-generated-content-in-political-advertisements}{URL} \\ (\href{https://perma.cc/HYH3-NEZV}{permalink})} \\
    \uid{EU PE 757.583} & \makecell[l]{Generative AI and Watermarking} & 2023 & \makecell[l]{European Parliament} & Briefing & \noTbl & \makecell[l]{\href{https://www.europarl.europa.eu/RegData/etudes/BRIE/2023/757583/EPRS_BRI(2023)757583_EN.pdf}{URL} \\ (\href{https://perma.cc/ALQ7-77DR}{permalink})} \\
    \uid{EU PE 751.478} & \makecell[l]{Artificial Intelligence, Democracy, \\ and Elections} & 2023 & \makecell[l]{European Parliament} & Briefing & \noTbl & \makecell[l]{\href{https://www.europarl.europa.eu/RegData/etudes/BRIE/2023/751478/EPRS_BRI(2023)751478_EN.pdf}{URL} \\ (\href{https://perma.cc/NV73-F59U}{permalink})} \\
     \bottomrule
\end{tabular}}
\caption{\textbf{List of other policy documents.}}
\label{fig:regulatory-docs-summary}
\label{tbl:regulatory-docs-summary}
\end{table*}

Now, we turn to other policy documents, with the goal of analyzing the relationship between these and legislative documents, highlighting points of synergy and disconnect.

\paragraph{Document overviews.} Our document selection collected 6 documents (4 US and 2 EU).
\iffullversion
    These comprised of guidance and briefings from various government agencies (NIST, CISA, FCC, and the European Parliament), as well as an executive order from the White House (that has since been revoked~\cite{ai-eo-revoked}\footnote{The White House recently released a new AI-related executive order~\cite{new-ai-eo}, but it neither contains watermarking-related mandates, nor is within the date range of our document inclusion criteria.}). 
    Table~\ref{fig:regulatory-docs-summary} lists these documents, including unique identifiers that we use for references within this Article\footnote{We use official document identifiers whenever one is available, and otherwise use our own identifiers.}, title, year of publication, drafting agency, document type, and whether it was been revoked or not.
    
\else
    Table~\ref{fig:regulatory-docs-summary} lists these documents and their details, including unique identifiers.
\fi
The other policy documents in our corpus serve diverse purposes, different also from the legislative corpus: 
from providing best practices (e.g., \uidreg{NIST AI 100-4}), to providing guidance for agencies and for future research (e.g., \uidreg{EO 14110}).
\iffullversion
    In contrast, the legislative documents generally aim to specify enforceable rules around AI content detection, in statutory language. 

\else 

\fi
Given the smaller number and diversity of focus within this document set, our analysis here is more flexibly structured than in Section \ref{sec:leg}. We begin with an overview of the six documents in our corpus, summarizing their content and purpose. We then present key top level themes from our inductive coding and thematic analysis. We connect these themes and documents to how they engage with the key trends and open questions identified in \secref{sec:leg}, and highlight two \textit{new} open questions exposed by these documents.

\subsection{Overview of Documents}

The White House Executive Order (\uidreg{EO 14110}) set up several directives for getting feedback or additional research from other agencies or regulatory bodies. This document provided explicit definitions of ``AI'', ``AI model'', and other relevant terms and sought to guide priorities in future study into relevant technologies. \uidreg{EO 14110} directed agencies to ``submit a report...identifying standards, tools, methods, and practices'' for authenticating, labeling, and detecting AI content. 

Guidance from NIST (\uidreg{NIST AI 100-4}) provided several details around implementing content transparency detection in response to the directives set forth by \uidreg{EO 14110}. It presented an extensive ``overview of technical approaches'' for tracking provenance and detecting AI generated content. 

Documents from FCC (\uidreg{FCC 2024}), CISA (\uidreg{CISA 2024}), and European Parliament (\uidreg{EU PE 757.583}, \uidreg{EU PE 751.478}) all focused on the context of elections and politics, and explored  tools for detecting  political AI content. 
\iffullversion
    \uidreg{FCC 2024}, from the FCC, focused on AI content in political advertising, and proposed a requirement for public broadcasters to disclose GenAI use, mirroring several bills in our corpus.
    This document requested additional input on policies it had already developed. 

    \uidreg{EU PE 751.478}, from the European Parliament, also described the pros and cons of AI related to political engagement, voting, and campaigns, and emphasized the threat of mis/disinformation. \uidreg{CISA 2024}, from CISA, provided an informational briefing on how AI content can be misused to impact elections, like through misinformation during elections or impersonating election officials. The document also provided several suggestions on how to protect against such harms. This focus mirrors \textbf{Key Trend~\ref{kt:onus-disclosure-users}}, showing heightened concern around AI content in election-related and political communication in both legislation and other forms of policy.

    \uidreg{EU PE 757.583}, also from the European Parliament, reviewed content transparency mechanisms, highlighting desirable traits such as the use of content transparency mechanisms in establishing content origin, while emphasizing open questions such as how to address false positives and robustness issues.
\fi

\subsection{Connection to Key Trends}\label{subsec:reg-takeaways}

Our goal here is primarily to highlight themes that emerged from our analysis and connect policy documents to the key trends of \secref{sec:leg}. We also identify additional open questions for content transparency policy. We organize our analysis by each broad theme identified, discussing relevant key trends and open questions within each section.

\paragraph{Technical scoping and limitations.} 
Our corpus of other policy documents vary widely in depth when discussing 
content transparency mechanisms
and their limitations. \uidreg{CISA 2024} does not name any content transparency mechanisms, and instead refers specifically to watermarks as a ``way to mark your content as verifiably originating from you...''. \uidreg{FCC 2024} is centered around self disclosure of AI use in political advertising, and focuses on how such disclosures should be communicated, without discussing how they may be removed .

The other documents (including all which are not specifically focused on the political communications context) each
provide far greater detail about content transparency mechanisms and their technical limitations than our legislative corpus did, including engagement with detection processes (striking a notable contrast with \textbf{Key Trend~\ref{kt:detection-process}}) and other technical properties and limitations of content transparency mechanisms (addressing limitations of the legislative documents noted in \textbf{Key Trends~\ref{kt:information}} and \textbf{~\ref{kt:properties}} and \textbf{Open Question~\ref{openq:reqs}}). 

\phantomsection\label{reftoappendix:regulatory-language-1}
For example, \uidreg{EO 14110} mentions watermarking specifically as a process for 
\iffullversion
    ``embedding information,
    which is typically difficult to remove, into outputs created by AI—including
    into outputs such as photos, videos, audio clips, or text—for the purposes
    of verifying the authenticity of the output or the identity or characteristics
    of its provenance, modifications, or conveyance.'' 
    This language implicitly requires some robustness, in mentioning the difficulty of removal.
\else
    ``embedding information,
    which is typically difficult to remove, into outputs created by AI \dots for the purposes
    of verifying the authenticity of the output or the identity or characteristics
    of its provenance, modifications, or conveyance.'' 
    This language implicitly requires some robustness.
\fi

\iffullversion
    \uidreg{EU PE 751.478} discusses post-hoc detection, invisible watermarks, metadata tagging, explicit labels, and visible watermarks. In doing so, the document highlights several drawbacks of the technologies themselves. When addressing post-hoc detection, the article explains that they ``can also easily be circumvented
by replacing a few words with synonyms, or by paraphrasing AI-generated text.'' When addressing invisible watermarks, the article points out that ``... it may reduce the quality of
AI-generated text due to its vocabulary constraint and that the watermarks can be removed by
paraphrasing...''

\uidreg{EU PE 757.583} provides a similar review of invisible watermarks, explicit labels, post-hoc detection, and visible watermarks, highlighting that post-hoc detectors can ``result in false positives'' and that invisible/visible watermarks ``can be manipulated, removed or altered (through backdoor attacks)'', and are ``vulnerable to spoofing
attacks, i.e. when an attacker (adversarial human) generates a non-AI text that is
detected as AI-generated.''. 
\uidreg{EU PE 757.583} also raises questions around ``deciding who should have the ability to detect the watermark signals, and decide whether the content is AI-generated and whether it is misleading...'', thus touching on the problem of determining violations with imperfect detection technology as discussed in \textbf{Key Trend~\ref{kt:violations}} and \textbf{Open Question~\ref{openq:imperfect-detection}}. This briefing recommends that detection ``could be made publicly available to allow users to ask whether an AI model had generated a specific item (wholly or in part).'' 

\iffullversion\else
    \uid{NIST AI 100-4} provides the most extensive discussion of limitations. 
    For example, the document lists five techniques for invisible watermarking, such as predictably modifying pixels or tokens in generated media, and presents their relative drawbacks. For predictable modification, this was ``[l]imited robustness and security (can be
    removed by, e.g., compression, cropping, filtering,
    scaling, or simple find/replace); perceptible distortions in the content; distortion and capacity may depend on the host content (e.g., texture of images)''. 
\fi

\fi

Standards and recommendations on how to implement technologies appear only in \uidreg{NIST AI 100-4}. The other policy documents 
provide limited commentary on how then to mitigate or manage the limitations they acknowledge. 
We emphasize that these documents and their authors have varied roles and are not directly comparable, and our comparisons and noting of omissions are descriptive rather than criticizing.
\uidreg{NIST AI 100-4} also provides the most extensive discussion of limitations. 
\iffullversion
    For example, the document lists five techniques for imperceptible watermarking, such as predictably modifying pixels or tokens in generated media, and presents their relative drawbacks. For predictable modification, this was ``[l]imited robustness and security (can be
    removed by, e.g., compression, cropping, filtering,
    scaling, or simple find/replace); perceptible distortions in the content; distortion and capacity may depend on the host content (e.g., texture of images)''. 
\else
    For example, the document lists five techniques for imperceptible watermarking.
\fi

\paragraph{Managing downstream harms.} These documents emphasize that watermarking and other content transparency mechanisms are not comprehensive solutions. \uidreg{CISA 2024}, for example, emphasizes that protecting against the harms of AI generated content go beyond showing just proof of origin or identifying something as generated using AI. Notably, the document discusses a series of \emph{downstream} harms from this AI content, such as to ``enable higher quality spearphishing attacks against election officials or staff to gain access to sensitive information.'' The document proceeds to recommend many non-detection related best practices to combat these, such as using multi-factor authentication (MFA), keeping social media private, and regularly requesting personal information be removed from public records. 

Similarly, \uidreg{EU PE 757.583} emphasizes how watermarking alone is insufficient for addressing the broader scope of AI harms, emphasizing that ``[i]t will have to be
accompanied by other measures, such as mandatory processes of documentation and transparency for foundation models, pre-release testing, third-party auditing, human rights impact assessments and media literacy campaigns.'' Other policy documents expose that holistically addressing harms from AI content is critical, and that recommendations of how to do so can often relate to broader security best practices. 

\paragraph{Allocating responsibility for compliance.}
While legislative documents tend to put the responsibility of disclosure on users (\textbf{Key Trend~\ref{kt:onus-disclosure-users}}), our corpus of documents pose responsibility allocation more as a complex unanswered question that requires further study, aligned with \textbf{Open Question~\ref{openq:responsibility}}.
They also discuss practical considerations about how compliance may work for different responsible parties, a topic that does not feature in legislative documents. 

\iffullversion
    For example, 
    \uid{FCC 2024} asks for comment on who should be responsible for disclosures in the context of third parties hosting content on broadcasting networks.
\begin{quote}
``Does the network or syndication company generally inform the broadcast stations airing the programming, at some point prior to the scheduled broadcast date, that a particular program includes a political ad?''
\end{quote} If so, the responsibility of disclosure would be on the network rather than the broadcasting station for disclosing political ads. The document also raises an alternate possibility:
\begin{quote}
``Should the
Commission require broadcast stations to make a simple inquiry to the respective network or syndication company, at the time the network or syndication company informs the stations airing the programming that a political ad is embedded in a particular program, whether the ad contains AI generated content?''
\end{quote}

\else
    \phantomsection\label{reftoappendix:FCC-broadcasting}
    For example, \uidreg{FCC 2024} asks for comment on who should be responsible for disclosures in the context of third parties hosting content on broadcasting networks, posing open questions and multiple alternative possibilities around how such responsibility would work in practice. 
\fi

\uidreg{EU PE 757.583} raises a specific nuance of responsibility allocation (\textbf{Open Question~\ref{openq:responsibility}}), also discussed in \secref{sec:leg}:
\begin{quote}
``[P]olicymakers and industry players will have to reflect
on how best to enforce watermarking in open-source ecosystems where different versions of open-source software can proliferate.''
\end{quote}

Furthermore, \uidreg{NIST AI 100-4} notes potential technical complications with regulating ``open'' GenAI models:
\begin{quote}
``Malicious actors generating this content 
\iffullversion
    on the Internet 
\else
    \dots
\fi
often use freely available models or build their
own smaller models based on existing open-source code, from
which they can easily remove safeguards.”
\end{quote} 

\iffullversion
    \uidreg{NIST AI 100-4} further describes  how ``open-source image generation packages include input data filters'' that filter out harmful requests, and that these are easier to bypass than filters with private models as they are publicly accessible.
\fi

Currently, there are no watermarks for open models that are not trivially removeable, and it is not clear whether this is even technically feasible at all to begin with~\cite{zhao2025watermarking}. Furthermore, such ease of circumvention connects to concerns of criminalizing commonplace or research activities, as discussed in \textbf{Open Question~\ref{openq:criminalizing}}.
Relatedly, \uidreg{EU PE 751.478} describes open-sourced models as a boon in that they ``would make it easier for the forensics community to build effective detection tools for tackling deepfakes, among
others''. 

\begin{openQuestion}
\label{openq:already-open-models}
    How would content transparency for open models work?
    How should content transparency regulation apply to open models? What about open models released prior to a regulation's enactment?
\end{openQuestion}

\iffullversion
    The document refers to the relative ease of developing robust solutions for an open model versus a proprietary one. This discussion acknowledges the limitations of current content transparency mechanisms, engaging with \textbf{Open Question~\ref{openq:imperfect-detection}}.
\fi

\paragraph{User experience and content transparency mechanisms.} 
Our corpus of other policy documents discuss user experience and usability of content transparency mechanisms, a topic absent in our legislative documents.

Understanding how users interact with and understand these mechanisms is essential for content transparency legislation's stated aim of mitigating the harms of AI-generated content that is indistinguishable from other content --- to the public, to consumers, and to the electorate. 

\uidreg{EU PE 757.583} discusses how ``typical notice and consent disclosures are largely ignored by users''. %
\iffullversion
    They put forward the intriguing suggestion that 
\else
    They suggest that
\fi
``the industry should develop AI language technologies that are self-disclosing by design (e.g. producing language that humans intuitively connect to AI sources and avoiding
language that people wrongly associate with humanity).''

\begin{openQuestion}
\label{openq:usability}
    How should content transparency mechanisms be designed for intuitive use and understanding by the public? How can policy initiatives encourage or mandate such usable design principles?
\end{openQuestion}

\iffullversion
    While little is known about the efficacy of content transparency mechanisms, regulatory documents highlight the need for careful study of disclosure design. 
\uidreg{NIST AI 100-4} makes this explicit, saying that ``there appears to be limited consensus from industry, civil society, and academia about how labels can be designed to best promote digital content transparency and how effective they can be. More research and evaluations are needed to inform effective
label design across context and use cases.'' The same document emphasizes  specific and understudied questions around usability.\footnote{For example, it correctly highlights that in scaling up signed metadata tagging,
``if content authors wish to verifiably sign as themselves, in addition to or instead of having tools or signing
services be the signers, the authors will need to apply for and manage their own keys and certificates, which
can be challenging for many users.'' Given that such models may be used for personal use, designing effective personal key management is critical and will require additional user study.} 

\else
    \phantomsection\label{reftoappendix:NIST-usability}
    Our corpus of documents also highlight the need for additional study of disclosure design, including aspects surrounding the \emph{usability} of disclosures.
    
\fi

\phantomsection\label{reftoappendix:tensions-existing-law}
\paragraph{Tensions with other existing law.} Our corpus of other policy documents discuss possible tensions between mandating content transparency mechanisms, free speech rights, and copyright laws. This is another topic that is absent from explicit mention in the legislative documents,\footnote{While not mentioned in the bill per se, \uidleg{S 97} in Louisiana is an exception, as it was vetoed by the Governor citing First Amendment concerns.} although behind the scenes, legislative provisions may well have been crafted with constitutionality and compatibility in mind.

\begin{openQuestion}
\label{openq:constitutional}
What approaches to content transparency policy respect freedom of speech, expression, and the press? How could legislation minimize potential constitutional challenges?
\end{openQuestion}

\iffullversion
    For example, \uidreg{FCC 2024} discusses the impact of deploying these technologies on free expression, asking whether requiring the disclosure of AI use in political ads ``violate the First Amendment rights of the candidates or other entities that sponsor political ads.'' \uidreg{FCC 2024} tentatively concludes instead that the proposed regulation ``would promote the goals of the First Amendment by enhancing the public’s ability to evaluate the substance and reliability of political ads, thus fostering an informed electorate.''

\uidreg{EU PE 757.583} discusses how techniques like invisible watermarking can better support other existing legal frameworks, since it ``would allow rights-holders to rely on the opt-out mechanism envisaged under the Copyright Directive.'' This was in referring to the EU Copyright Directive and its mandate to allow creators and copyright owners to opt-out of including their media in model training data. \uidreg{EU PE 751.478} describes how these technologies may fit with existing EU legislation, such as the AI Act. 

\fi

\section{Critical Areas for Future Content Transparency Policy}\label{sec:recommendations}
Our analysis thus far makes clear that many questions remain unresolved, with critical points of disconnect between content transparency policy and technical theory and practice. Next, we synthesize broad recommendations and critical areas for future development in the security, AI, and policy communities. Our discussion here is purposefully general: our goal is not to say how policy should be written, but rather to point towards broad areas of importance that remain, thus far, under-studied.

\iflaw
\else
\paragraph{Requirements should be context-sensitive.}
\fi
Detection of AI-generated content across different contexts entails different considerations and constraints, as evidenced by some of the differences between legislative documents focused on the election context. Blanket regulation across all contexts may bring higher risk of conflict with freedom of speech (\textbf{Open Question~\ref{openq:constitutional}}), as well as higher risk of criminalizing commonplace or research activity (\textbf{Open Question~\ref{openq:criminalizing}}). Further study of contextual differences is essential to tailor regulation to avoid overbreadth, unintended side effects, and legal challenges.

\iflaw
\else
\paragraph{User privacy is under-studied.} 
\fi
User privacy considerations do not feature prominently in our corpus, yet content transparency raises privacy issues as it enables tracking of user expression and may involve embedding sensitive (personal) information in disclosures. Privacy considerations are moreover related to freedom of speech, expression, and political beliefs~\cite{un-anon-report}.
Content transparency could facilitate leakage of sensitive information; not only could this harm end users, but it could also result in tensions with existing data protection frameworks --- 
\iffullversion
    a type of tension not mentioned in our corpus despite the discussion of possible conflicts with existing laws in some of our corpus of other policy documents (discussed in \secref{sec:reg}).
\else
    a type of tension not mentioned in our corpus.
\fi

\iflaw
\else
\paragraph{Technical properties involve nuanced trade-offs, which should not be left to technologists alone.} 
\fi
The precise technical properties of content transparency systems have significant impacts on their fitness for different applications. As noted above, improving one property may limit another, and it is often necessary to make tradeoffs between different properties. 
Yet, as noted in \textbf{Key Trends~\ref{kt:information}}, \textbf{\ref{kt:detection-process}}, we have found that policy discourse engages rarely with this level of technical detail  --- effectively, not differentiating between different schemes serving very different purposes.

These are not implementation details that can be left to technologists' engineering expertise; rather, policy instruments should provide more actionable guidance on appropriate trade-offs to strike for different policy goals. This may require coordination between policy and technology experts.

For example, one might accuse a model provider of generating an offensive image, citing as evidence the fact that the image carries that provider's watermark.
However, the vast majority of content transparency mechanisms are \emph{not} fit for this application, as they are \emph{forgeable}.
That is, given enough content with disclosures, an attacker can learn to embed that disclosure in content of its choice.
The desired property in this scenario is typically called \emph{unforgeability}.

As another example, a model provider might attempt to prove that it generated an image, by showing that the image carries its watermark.
Most watermarks are not fit for this use either, as it is possible to take any fixed image and manufacture a watermarking key that is consistent with it.

We urge deeper policy engagement with technical properties of content transparency mechanisms as an essential step towards coherent efforts towards specific policy objectives---for general content transparency legislation, as well as context-specific legislation motivated by specific applications such as the examples above.

\iflaw
\else
\paragraph{Consider who we (need to) rely on for infrastructure.}
\fi
Notably entirely absent from our corpus, though highlighted in \textbf{Open Question~\ref{openq:trust}},
is discussion of which parties are or should be relied upon for content transparency infrastructure to function correctly. Which aspects require the model provider's cooperation (if any)? Who is responsible for system design? What about implementation?

Many existing approaches require model providers' cooperation in order to embed watermarks.
This means the model provider can decide whether to watermark each output.
Furthermore, in many schemes, detection requires a secret key.
A natural way to enable the public to detect the watermark is for the party holding this key to offer a detection API.
However, this requires significant trust in the API operator which could, for example, provide incorrect responses. Would this be detectable? How?
Who should the API operator be, and what is the rationale for entrusting them with this role?

Such factors importantly influence the trustworthiness and possible failure modes of deployed content transparency mechanisms. As such, we argue that policy discourse should include explicit discussion of the entities to be relied upon for content transparency infrastructure,  their powers, obligations, and incentives, what could go wrong if these entities were compromised, and possible alternative arrangements.

\iflaw
\else
\paragraph{Criminalizing disclosure removal raises serious risks.}
\fi
As emphasized in \textbf{Open Question~\ref{openq:criminalizing}}, while criminalizing the removal of disclosures appears as a common theme in our legislative documents, such provisions raise serious risks of criminalizing a broad swath of commonplace activity, and of criminalizing research essential to our understanding and improvement of content transparency mechanisms. Moreover, differentiating between benign and deliberate removals is challenging due to the fact that disclosures tend to have low robustness, and can be removed by commonplace operations such as text translation and whitespace normalization. 

\iffullversion
    Similarly, overbroad computer crime laws are known to create serious legal risks around commonplace activities such as scraping and reverse engineering (and more), with negative repercussions, including chilling effects on important research that would improve security online \cite{PA24,GSLS17,do-the-right-thing}.
    Taking heed of decades of experience in this similar context, we have serious concerns about criminalizing disclosure removal, and stress the importance of clear and carefully designed limits on the scope of any criminalization.
\else
    Similarly, overbroad computer crime laws have long been known to create serious legal risks around commonplace activities such as scraping and reverse engineering (and more) \cite{PA24,GSLS17,do-the-right-thing}. %
    We urge caution and stress the importance of clear limits on criminalization.
\fi

\iflaw
\else
\paragraph{Interrogate how content transparency will help.}
\fi
We have seen existing discourse exhibiting a strong focus on content transparency as a solution or mitigation to various kinds of harms, such as related to mis- and disinformation and other examples noted in \secref{sec:intro}. Yet it does not engage in depth with exactly how content transparency will address the problems that motivate the legislation or policy at hand---especially considering that most of the motivating harms are ones that humans can cause and have long caused without AI: for example, (politically motivated) mis- and disinformation, scams, fraud, plagiarism, and more.\iflaw\else\footnote{That is, essentially all the examples from the beginning of \secref{sec:intro}, except for recursive model training, which we believe is unique to AI but is not an apparent motivation for any of the documents we reviewed. }\fi

Deeper engagement with exactly how content transparency is anticipated to help with specific problems in concrete application contexts will facilitate the selection of appropriately configured tools and the development of new tools that are a better fit---thus making progress on \textbf{Open Questions~\ref{openq:disclosure-types} and \ref{openq:reqs}}---as well as support determinations that content transparency is \emph{not} the right fit as the case may be.

Finally, we highlight the broader questions of (1) how, if at all, AI-generated content truly changes the nature of and mitigations for mis- and disinformation, fraud, scams, plagiarism, and other harms commonly associated with AI, and (2) relatedly, whether and when distinguishing of AI-generated content from human-authored content is likely to help. While a thorough examination of these questions is outside the scope of the present Article, we believe that they are essential to the future of content transparency policy, and present a fruitful and timely area for future work.

\section{Further Related Work}\label{sec:related}

\iffullversion
    We briefly highlight additional areas of related work.
\else
    This paper does not systematize prior academic research, and we do not view our work as a systematization of knowledge (SoK) paper (see Appendix~\ref{sec:A-why-not-sok} for further discussion). However, we briefly highlight a number of related work.
\fi

\paragraph{AI and policy.} Our paper complements broader emerging literature examining GenAI and policy~\cite{christodorescu2024securing,nemecek2025watermarking,rijsbosch2025adoption,hacker2023regulating}. 

For example, Christodorescu et al.~\cite{christodorescu2024securing} is a broad summary of a workshop on GenAI policy, which mentions content transparency, and generally calls for more alignment between watermarking policy and technical capabilities.
Hacker et al.~\cite{hacker2023regulating} offer a broad critique of existing GenAI regulation and some suggestions. Their work only covers the EU AI Act and Digital Services Act in detail, and only mentions watermarking briefly.
Nemecek et al.~\cite{nemecek2025watermarking} and Rijsbosch~\cite{rijsbosch2025adoption} focus more directly on content transparency. 
The primary contribution of~\cite{nemecek2025watermarking} is an abstract framework for future content transparency policy; complementarily, we analyze existing policy efforts.
\iffullversion
\footnote{To support their framework,~\cite{nemecek2025watermarking} briefly mentions some existing policy documents, and quotes from five bills; but this is not their focus and their review of existing policy is not systematic.}
\else
\fi
While~\cite{rijsbosch2025adoption} does discuss existing legislation, they focus on the EU AI Act.

\paragraph{Content transparency in computer science.} On the technical side, there is an extensive literature on the development and analysis of content transparency mechanisms, spanning security, cryptography, and machine learning. These include proposals for new definitions and security models~\cite{DBLP:conf/colt/ChristGZ24,DBLP:journals/tmlr/KuditipudiTHL24,DBLP:journals/cic/FairozeGJMMW24}, constructions of new schemes~\cite{DBLP:conf/icml/KirchenbauerGWK23,scott,DBLP:journals/nature/DathathriSGHMWBKSMHVMBB24}, security analyses of existing technologies~\cite{DBLP:conf/icml/0001SV24,DBLP:conf/nips/ZhaoZSVGKVWL24}, and more.

Invisible watermarks are a notable sub-area, and many relevant works have appeared at security (S\&P, USENIX, CCS \cite{DBLP:conf/sp/Cohen0S25,DBLP:conf/uss/ZhangHNK24,DBLP:conf/ccs/JiangZG23}) and cryptography (CRYPTO, CiC \cite{DBLP:conf/crypto/ChristG24,DBLP:journals/cic/FairozeGJMMW24}) venues. We refer to Zhao et al.'s systematization of knowledge on invisible watermarking~\cite{zhao2025watermarking} (S\&P 2025) for a recent overview and extensive references. Most of the invisible watermarking literature is purely technical, covering formal definitions of security properties, threat models, evaluation methodologies, and existing constructions.

\paragraph{AI and democracy.} A number of works study the role of GenAI in democratic processes~\cite{chiacchiaro2025generative,persily2025misunderstanding,allen2024real},
where watermarking is often referenced as potentially useful to enhance election integrity and/or mitigate mis- and dis-information. 
Notably,~\cite{chiacchiaro2025generative}
does engage with existing policy developments, but in a much more limited scope than us (mostly focused on electoral communications) and from a purely legal angle (in particular, without much consideration of the technical aspects of content transparency).

\section{Conclusion}
\label{sec:conclusion}
\iffullversion
    The rapid advancement of generative AI models brings both exciting applications and significant sociotechnical challenges.
    As models increase in sophistication and their outputs become more realistic,
\else
    As model outputs become more sophisticated and realistic,
\fi
distinguishing AI- from human-generated content becomes more pressing. Content transparency and ``AI watermarking'' are prominently discussed approaches to mitigating some potential harms, in both technology and policy circles. Yet discourse on these topics has been fragmented across and within disciplines, and sometimes policy language can appear disconnected from technical practice.%

This Article provides the first systematic analysis of discourse on content transparency and ``AI watermarking'' policy, with a deeply interdisciplinary lens combining expertise in technology law and content transparency technologies. We analyze existing developments, extract key trends, and identify open questions, ambiguities, and gaps between legislative proposals and technical practice. We offer five concrete recommendations, and provide common ground for the structured, cross-disciplinary analysis of content transparency policy
that will be much needed as the field matures.

    \section*{Acknowledgments}
    We are grateful to Sam Gunn for extensive discussions in early stages of the project, to Stephanie Chen for assistance with background research, and to Jesse Dunietz, Jason Schultz, and James Grimmelmann for helpful feedback on our ideas and draft versions.
    
    Miranda Christ was partially supported by a Google Cyber NYC grant, an Amazon Research Award, and NSF grants CCF-2312242, CCF-2107187, and CCF-2212233. Sunoo Park was partially supported by the Institute of Information \& Communications Technology Planning \& Evaluation (IITP) with a grant funded by the Ministry of Science and ICT (MSIT) of the Republic of Korea, the Global AI Frontier Lab International Collaborative Research (No. RS-2024-00469482 \& RS-2024-00509258).

    \bibliographystyle{plain}
    \bibliography{references}

\appendix
\newpage
\section{Recap of Key Takeaways and Open Questions}\label{sec:kt-oq}
In this section, we list all key trends and open questions identified in
\secref{sec:leg,sec:reg}.
They are exactly the same as they appeared earlier in the Article, simply collected in one place for easy reference.

\subsection{Key trends}
\setcounter{keyTrendEnv}{0}
\begin{takeaway}
    Legislative documents tend to leave considerable room for interpretation in their scope of covered content. 
\end{takeaway}

\begin{takeaway}
    Bills that narrow their scope to content that is ``substantially modified'' by AI often do so in terms of changes to the content's perceived meaning or truth provenance.
\end{takeaway}

\begin{takeaway}
    A majority of legislative documents place the onus of disclosure on the users of AI tools.
\end{takeaway}

\begin{takeaway}
    In contrast to computer-science literature, which generally focuses on properties of \emph{models}, bills often impose blanket requirements on \emph{model outputs}. This discrepancy can make proposed requirements ambiguous or technically unachievable when literally interpreted.

\end{takeaway}

\begin{takeaway}
    Many bills prohibit the removal of disclosures, which could criminalize commonplace activities and essential content transparency research.
\end{takeaway}

\begin{takeaway}
    ``Watermark'' is often used ambiguously within and across bills. Disclosure requirements vary considerably, and often are not clear cut.    
\end{takeaway}

\begin{takeaway}
    Many bills require that specific information is included in disclosures, such as model information or creation metadata. This may reveal sensitive information, or constrain the technologies that may be used to satisfy the bill's requirements.
\end{takeaway}

\begin{takeaway}\label{kt:detection-process}
    Most bills do not acknowledge detection processes for imperceptible disclosure mechanisms.
\end{takeaway}

\begin{takeaway}\label{kt:properties}
    Bills do not engage in depth with essential security properties of content transparency mechanisms (such as robustness) and the tradeoffs between them.
\end{takeaway}

\begin{takeaway}\label{kt:violations}
    It is unclear how violations can and will be determined, and how enforcement will account for the limitations of content transparency mechanisms. 
\end{takeaway}

\subsection{Open questions}
\setcounter{openQuestionEnv}{0}
\begin{openQuestion}
    How should we determine what constitutes `AI-generated' content? To what extent should such determinations be context-dependent?
\end{openQuestion}

\begin{openQuestion}
    Who should bear the responsibility of disclosure in which contexts, and how should responsibility be divided between parties?
\end{openQuestion}

\begin{openQuestion}
    Who are we relying on for content transparency infrastructure?
\end{openQuestion}

\begin{openQuestion}
    Can criminalization of disclosure removal avoid criminalizing commonplace activities and chilling essential content transparency research?
\end{openQuestion}

\begin{openQuestion}
    Which content transparency mechanisms are best suited to achieve the goals of disclosure in different contexts? How should legislative documents specify these clearly while maintaining flexibility to adapt to future technological change?
\end{openQuestion}

\begin{openQuestion}\label{openq:reqs}
    What technical requirements are most relevant to content transparency policy, and how should legislative documents acknowledge them?
\end{openQuestion}

\begin{openQuestion}
\label{openq:imperfect-detection}
    How can violations be determined reliably while taking into account the limitations of detection and labeling technologies?
\end{openQuestion}

\begin{openQuestion}
\label{openq:already-open-models}
    How would content transparency for open models work?
    How should content transparency regulation apply to open models? What about open models released prior to a regulation's enactment?
\end{openQuestion}

\begin{openQuestion}
\label{openq:usability}
    How should content transparency mechanisms be designed for intuitive use and understanding by the public? How can policy initiatives encourage or mandate such usable design principles?
\end{openQuestion}

\begin{openQuestion}
\label{openq:constitutional}
What approaches to content transparency policy respect freedom of speech, expression, and the press? How could legislation minimize potential constitutional challenges?
\end{openQuestion}

\end{document}